\documentclass[%
 aip,
 amsmath,amssymb,
preprint,%
floatfix
]{revtex4-1}

\usepackage{graphicx}
\usepackage{dcolumn}
\usepackage{bm}

\usepackage[utf8]{inputenc}
\usepackage[T1]{fontenc}
\usepackage{mathptmx}

\begin{document}

\preprint{}

\title[]{Strain-controlled Domain Wall injection into nanowires for sensor applications}

\author{Giovanni Masciocchi}
\email{gmascioc@uni-mainz.de}
 \affiliation{Institute of Physics, Johannes Gutenberg -
University Mainz, 55099 Mainz, Germany.}
 \affiliation{Sensitec GmbH, 55130 Mainz, Germany.}
\author{Mouad Fattouhi}%
\affiliation{ 
Universidad de Salamanca, Department of Applied Physics, E-37008 Salamanca, Spain 
}%

\author{Andreas Kehlberger}
\affiliation{Sensitec GmbH, 55130 Mainz, Germany.}
\author{Luis Lopez-Diaz}%
\affiliation{ 
Universidad de Salamanca, Department of Applied Physics, E-37008 Salamanca, Spain 
}%
\author{Maria-Andromachi Syskaki}
\affiliation{Institute of Physics, Johannes Gutenberg -
University Mainz, 55099 Mainz, Germany.}
\affiliation{Singulus Technologies AG, 63796 Kahl am Main, Germany.}
\author{Mathias Kläui}
\affiliation{Institute of Physics, Johannes Gutenberg -
University Mainz, 55099 Mainz, Germany.}

\date{\today}

\begin{abstract}

 We investigate experimentally the effects of externally applied strain on the injection of 180$^\circ$ domain walls (DW) from a nucleation pad into magnetic nanowires, as typically used for DW-based sensors. In our study the strain, generated by substrate bending, induces in the material a uniaxial anisotropy due to magnetoelastic coupling. To compare the strain effects, $Co_{40}Fe_{40}B_{20}$, $Ni$ and $Ni_{82}Fe_{18}$ samples with in-plane magnetization and different magnetoelastic coupling are deposited. In these samples, we measure the magnetic field required for the injection of a DW, by imaging differential contrast in a magneto-optical Kerr microscope. We find that strain increases the DW injection field, however, the switching mechanism depends strongly on the direction of the strain with respect to the wire axis. We observe that low magnetic anisotropy facilitates the creation of a domain wall at the junction between the pad and the wire, whereas a strain-induced magnetic easy axis significantly increases the coercive field of the nucleation pad. Additionally, we find that the effects of mechanical strain can be counteracted by a  magnetic uniaxial anisotropy  perpendicular to the strain-induced easy axis. In $Co_{40}Fe_{40}B_{20}$, we show that this anisotropy can be induced by annealing in a magnetic field. We perform micromagnetic simulations to support the interpretation of our experimental findings. Our simulations show that the above described observations can be explained by the effective anisotropy in the device. The anisotropy influences the switching mechanism in the nucleation pad as well as the pinning of the DW at the wire entrance. As the DW injection is a key operation for sensor performances, the observations show that strain is imposing a lower limit for the sensor field operating window.

\end{abstract}

\maketitle



\section{Introduction}
Domain walls (DWs) have always been of importance for their static and dynamic properties since the use of magnetic materials for logic devices and data storage\cite{klaui2008head,1065929}. In the last 15 years, the possibility to realize and characterize magnetic nanostructures has allowed one to explore complex spin textures, their creation and stability. This intense research has enabled the use of DWs in memory devices\cite{parkin2008magnetic} and magnetic sensors\cite{jogschies2015recent}. For example, a DW can be used to carry information about the angular position of an object and to count the number of rotations performed in a non-volatile way\cite{diegel2007multiturn,mattheis2012concepts, borie2017geometrically}. The interest of magnetic sensors based on DWs is in their stability, making their non-volatile positioning suitable to many applications. No external electrical power is required to manipulate the magnetic state in the sensor, making it ideal for energy efficient systems even where power failures can occur. 

The magnetic field conditions under which a DW based sensor can reliably operate are called field operating window\cite{borie2017geometrical}. For the sensor to work, a DW needs to be successfully created and propagated into the nanowire, setting the minimum operation field value. At the same time, uncontrolled nucleation of domain walls at higher fields needs to be avoided, thus setting the maximum operation field value. Previous studies about  DW sensors investigated the propagation and nucleation fields, and showed how they depend on material parameters and device geometry \cite{martinez2007thermal, martinez2009domain, garcia2011depinning,borie2017geometrical,borie2017geometrically,hoang2020creation}. While the field operation window in idealized operation conditions is known, in real devices further factors play a role and have been previously neglected.

\begin{figure}[h!]
\centering\includegraphics[width=7cm]{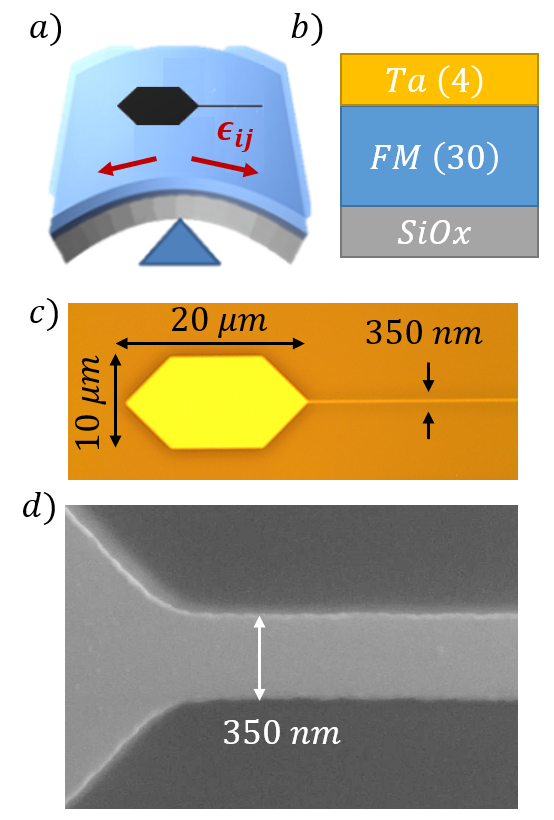}
\caption{\label{fig_structure} a) Schematic of the mechanism to apply mechanical strain by substrate bending. b) Layer cross-section used for the investigated devices. FM indicates the magnetic material, the numbers correspond to the thickness in $nm$. The $SiOx$ thickness is $1.5$ $ \mu m$.  Device shown in an optical microscope c)  and scanning electron microscope image d). }
\end{figure}

Among the external factors, strain  or mechanical stress on these sensing elements is known  to be a key issue. Such strain occurs during packaging as well as sensor operation, with strong impact on the device performance\cite{van2003packaging}. Strain in magnetic materials is known to induce a preferential direction of magnetization (anisotropy) due to magnetoelastic coupling \cite{lee1955magnetostriction,finizio2014magnetic} and even pin a DW in a nanowire\cite{lei2013strain}.  In DW based devices, a common approach to generate a DW is to use a larger magnetic (nucleation) pad attached to the nanowire exploiting the reduced shape anisotropy\cite{cowburn2002domain, shigeto1999injection,hoang2020creation} as shown in figure \ref{fig_structure}. It has been shown recently using simulations how, in the nucleation pad, strain-induced anisotropy can overcome the shape anisotropy governing the switching of the magnetic state  \cite{zhou2020voltage}. However, these previous studies did not report experiments on strain effects in a sensor relevant system as ours. 

In the work presented here, we investigate experimentally the impact of externally applied strain on the injection of a 180$^{\circ}$ domain wall from a nucleation pad in a  magnetic nanowire. We employ magneto-optical Kerr effect (MOKE)  microscopy to image the DW creation, pinning and injection from the pad for different external strain configurations. The injection field is strongly affected by the effective anisotropy of the magnetic material, which is modified by strain. Simulations are used to identify the switching mechanism and the spin structure of a pinned DW  just before the injection into the wire.


\section{Experimental methods}

We investigate three different samples:  $Co_{40}Fe_{40}B_{20} (30$ $nm)/Ta(4$ $nm)$, $NiFe_{11}Cr_{42}(4$ $nm)/$ $Ni_{82}Fe_{18}(30$ $nm)/Ta(4$ $nm)$ and  $Ni(30$ $nm)/Pt(2 $ $nm)$, all layers are deposited by magnetron sputtering. The substrate is $1.5$ $\mu m$ thick, thermally oxidized $SiOx$ on top of 625 $\mu m$ undoped $Si$. To improve amorphisation  and deposition uniformity, the magnetic material was sputtered in a rotating magnetic field of $50$ $Oe$. The result is a soft magnetic material with intrinsically low anisotropy field and low coercive field. The magnetic properties of our films, as deposited, are summarized in table \ref{tab_material_film}. For the characterization of our material we used a BH-Looper, an  hysteresis loop tracer, where B is the measured magnetic flux and H is the applied magnetic field (Shb Instruments - Model 109). This tool includes a setup for measuring magnetostriction\cite{choe1999high,hill2013whole,raghunathan2009comparison}.

\begin{table}[h!]
    \centering
    \begin{tabular}{||c c c c c||} 
 \hline
    Material & $M_{s}$ $ (T)$ & $B_k$ $(mT)$  &  $B_c$ $(mT)$  & $\lambda_s$ x$10^{-6}$ \\ [0.5ex] 
 \hline\hline
 $Co_{40}Fe_{40}B_{20}$ & 1.40(5) &	0.20(5) &	0.10(5) &	27(1) \\
  \hline
 $Ni_{82} Fe_{18}$ & 0.95(5) &	0.10(5) &	0.10(5) &	-0.5(1) \\ 
 \hline
 $Ni $ &  0.60(5)  &	2.00(5) &	2.00(5) &	-32(1)\cite{cullity2011introduction} \\ 
 \hline
\end{tabular}
\caption{Parameters of the magnetic materials (thickness $30$ $nm$) after deposition (no annealing). The experimental values are obtained by measuring the magnetic film on $5"$ wafers. Here, $M_s$ is the saturation magnetization, $B_k$ is the anisotropy field, $B_c$ is the coercive field and $\lambda_s$ is the saturation magnetostriction.}
\label{tab_material_film}
\end{table}

\begin{figure}[h!]
\centering\includegraphics[width=12cm]{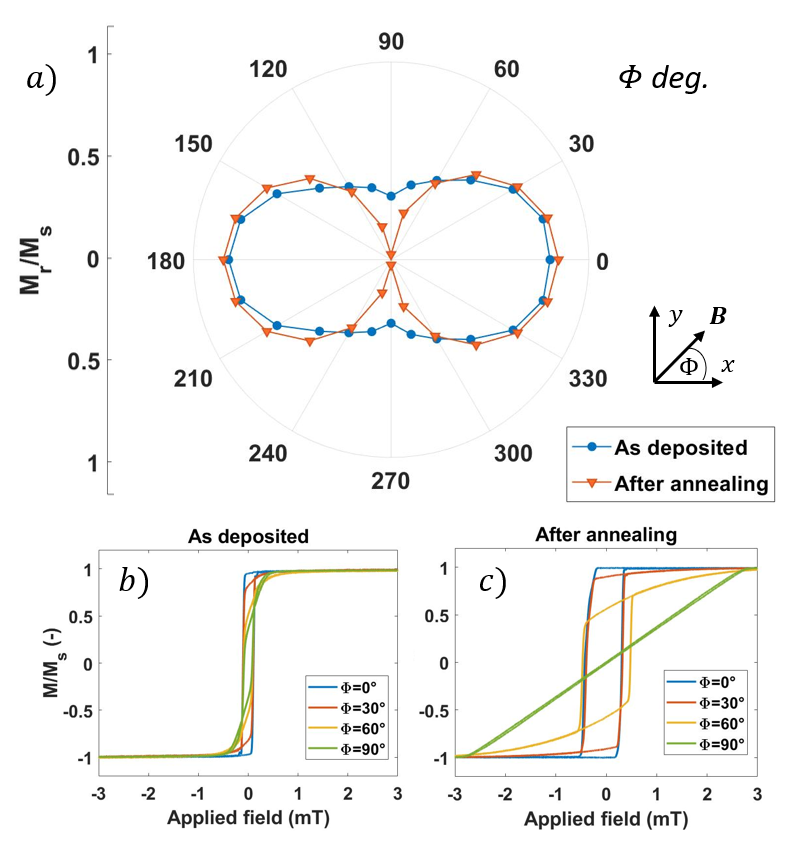}
\caption{\label{fig_loops}  Characterization of the of the full film 5” wafers of $Co_{40}Fe_{40}B_{20}$ using a BH Looper before structuring.  a) The angular dependence of the remanent magnetisation $M_r/M_s$ shows the effects of thermal annealing in presence of magnetic field (orange triangles), that induces a uniaxial anisotropy with easy axis in the direction $\Phi= 0^\circ$. Error bars are within the data points. The magnetic curves of the sample as deposited b) and after annealing c) show a easy axis and a hard axis of magnetization in green and blue, respectively. Orange and yellow are intermediate direction of applied magnetic field.}
\end{figure}

To induce a preferential direction of magnetization in the $Co_{40}Fe_{40}B_{20}$, the sample was annealed in $N_2$ for 2 hours at $T=265^{\circ}C$, while a static field of $120$ $mT$ is applied. This treatment induces a uniaxial anisotropy\cite{zhang2011study}, where the easy axis of the magnetization is in the direction of the applied magnetic field. The anisotropy field after annealing is $B_k=\frac{2 K_{film}}{\mu_0 M_s}=2.7$ $mT$ and the uniaxial anisotropy constant is $K_{film}= 1.54(2)$ $\frac{kJ}{m^3} $, where $M_s$ is the saturation magnetization of the magnetic material. A comparison of the angular dependence of the remanent magnetization and hysteresis loops before and after annealing are shown for $Co_{40}Fe_{40}B_{20}$, respectively in figure \ref{fig_loops} a) and b) - c).
The structures are then patterned using photolithograpy and Ar ion milling.


The devices used in this study can be seen in figure \ref{fig_structure} c). A nucleation pad ($20$ $\mu m \times 10$ $ \mu m $) is attached to a $200 $ $\mu m$ long nanowire with different widths, from $350$ $ nm$ to $800$ $ nm$. The specific geometry of the pad is designed to narrow the field distribution for injecting magnetic domains into the wire, and to allow for a DW nucleation at low fields.

 To  switch the magnetization in the device, we applied an external in-plane magnetic field in the $x$ direction (aligned with the nanowire). As the magnitude of the field is increased, the nucleated DW depins from the pad and is injected into the nanowire (injection field, $B_{inj}$). We measure the injection field by imaging differential contrast changes in the magneto-optical Kerr effect (MOKE)
in a longitudinal configuration of the polarized white light.  To image and detect the switching event, a 50x magnification objective was used. The magnetic contrast, without structural
contrast, is accomplished by subtracting a reference image in the saturated stage, at the beginning of the measurement. We have conducted our experiment at fields lower than the spontaneous domain nucleation field in the wire, that is reported to be around $40 $ $mT$ \cite{borie2017geometrical}. This ensures us that a DW is injected from the pad into the wire, and not from structure defects or nucleated at the edge of the wire. 

To apply strain to our devices, the substrate was bent mechanically  with a sample holder that applies an out of plane force as shown schematically in figure \ref{fig_structure} a). A square sample of 1 by 1 cm is clamped on two sides and pushed by a cylinder uniformly from below. The device generates a tensile strain in the plane of the sample up to $0.12\%$. The strain is mostly uniaxial\cite{raghunathan2009comparison} and uniform in the central area of the sample, and thus in the measured area. The intensity of the strain induced on the surface of the $SiOx$ has been measured with a strain gauge (RS PRO). The stack is thin enough to assume that the strain is entirely transferred to our device and that shear strain is negligible\cite{thomas2003interplay}.

\section{Results and discussion}

\subsection{Injection field in nanowires}

The injection field is the minimum field required to create and propagate a DW in the magnetic sensor device. However, to obtain reliable and a repeatable injection, one needs to understand the whole process of the DW injection. The MOKE images allow us to observe  how the magnetization is gradually switched in the nucleation pad, creating a DW at its end, and also, captures the moment when a DW starts propagating into the wire.


\begin{figure*}
\centering\includegraphics[width=15cm]{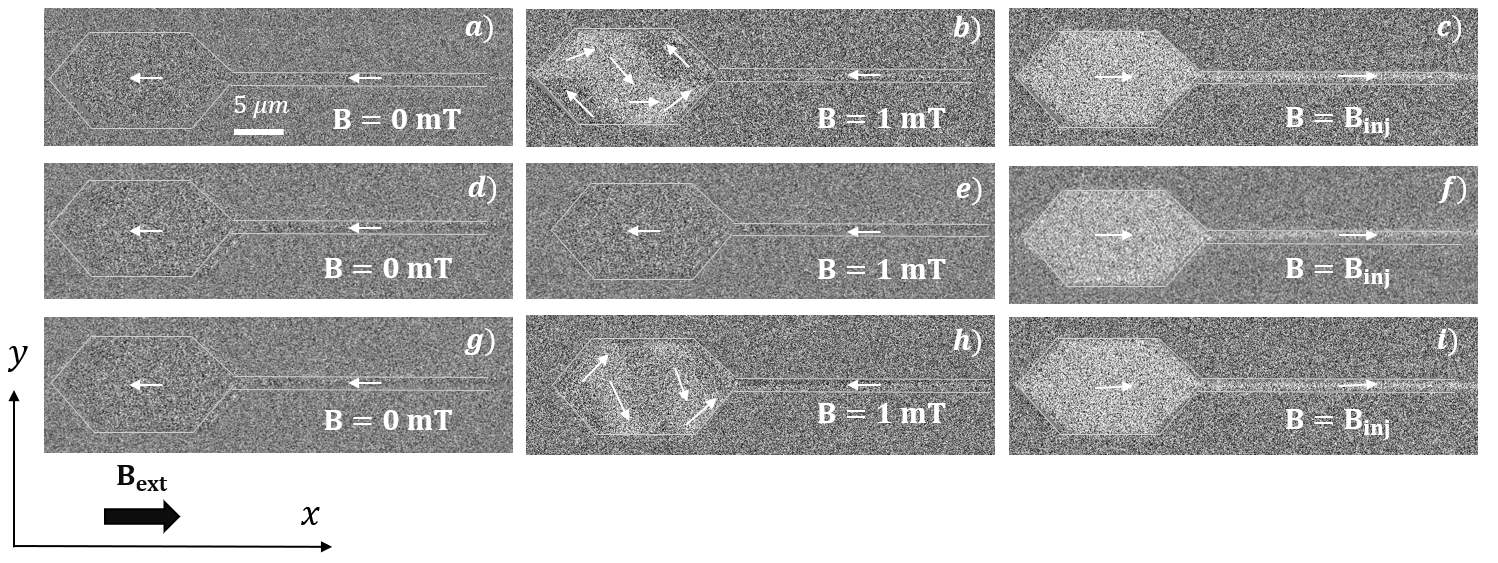}
\caption{\label{fig_switching}  Kerr microscope images of the device made of $Co_{40}Fe_{40}B_{20}$ as deposited (no annealing). The white arrows indicate the local direction of the magnetization. The width of the nanowire is $350$  $nm$. The field is applied along the $x$ direction and progressively increased (from left to right) until the DW nucleated in the pad is injected into the nanowire. In a) - c) the sample is not strained, d) - f) tensile strain is applied along the wire $\epsilon_{xx}=0.06\%$, and g) - i) tensile strain is applied perpendicular to the wire $\epsilon_{yy}=0.06\%$.}
\end{figure*}


In the absence of strain, the domain configuration in the pad is mostly dominated by the shape anisotropy, and is not varying strongly for the materials studied. We indeed obtain similar images for all the measured devices. The process of injection can be observed in figure \ref{fig_switching} a) - c). When no strain is applied to the substrate, a pattern with 6 domains is formed, with a domain wall at the entrance of the nanowire for fields as small as $1$ $ mT$. However, at such low fields the DW cannot propagate into the nanowire, and is pinned at its entrance. The difference in geometry between the pad and the wire creates a local pinning site for the DW. The pad has a lower coercivity than the nanowire due to the shape anisotropy, therefore, changes in the magnetization distribution of the pad result in the creation of a wall in the vicinity of the pad/wire interface. It is this wall that can then subsequently be injected to propagate along the wire\cite{mcgrouther2007controlled}. To obtain the injection, the external field is increased.

If, on the other hand, the substrate of our device is mechanically deformed (strained), an additional anisotropy is induced in the magnetic system. Strain-induced anisotropy will compete with the shape anisotropy and exchange to determine the domain configuration and the switching mechanism in the injection pad. It is known how the magnetization is coupled to the uniform macroscopic strain in the expression of the free energy\cite{wang2019mechanical}.  As reported in previous studies \cite{finizio2014magnetic,bur2011strain} the magnetoelastic energy simplifies to a uniaxial magnetic anisotropy constant defined as

\begin{equation} \label{eq_K_ME}
K_{ME}=\frac{3}{2}\lambda_s Y |\epsilon_{xx}-\epsilon_{yy}|
\end{equation}

where $\lambda_s$ is the saturation magnetostriction, $Y$ is the Young's modulus of the ferromagnetic layer and $\epsilon_{xx}$, $\epsilon_{yy}$ are the components of the uniaxial in-plane strain along, respectively, x and y. In our experiments, the magnitude of the strain is equal to $\epsilon_{ii}=0.06\%$, where $ii$ indicates the direction of uniaxial strain. Since the strain is uniaxial, we assume that the other direction can be neglected ($\epsilon_{yy}<<\epsilon_{xx}$ and vice versa). This means the strength and the direction of the uniaxial anisotropy contribution will be determined by, respectively, the magnitude and the sign of the saturation magnetostriction $\lambda_s$. In a positive magnetostrictive material ($Co_{40}Fe_{40}B_{20}$), the easy axis will follow the direction of the tensile strain, while there will be  a hard axis in this direction for a negative magnetostrictive material ($Ni$).

We experimentally observe that the conditions for the DW nucleation and injection are modified by the strain. When the easy axis of magnetization is along the wire ($x$ - direction), the coercive field of the pad is increased and the magnetization rotates suddenly from left to right. This can be seen in figure \ref{fig_switching} d) - f). No intermediate multi-domain state is present in the pad and the DW is not pinned at wire entrance, since the coercive field coincides with the injection field. The second case is represented in figure \ref{fig_switching} g) - i). Here the easy axis of magnetization is induced perpendicular to the field direction (along $y$). In this case the multi-domain state that minimizes the energy in the pad prefers spin aligned along the y axis, and a DW is nucleated at the entrance of the pad. In this case, the nucleated DW stays pinned at the entrance even for higher field with respect to the unstrained  case.
 
Using the values in table \ref{tab_material_film} and equation \ref{eq_K_ME}, one can calculate the uniaxial anisotropy constant  due to the magnetoelastic term which is  $K_{ME}=3.6(1)$ $\frac{kJ}{m^3}$ for $Co_{40}Fe_{40}B_{20}$,  $K_{ME}=-4.3(1)$ $\frac{kJ}{m^3}$ for $Ni$ and only  $K_{ME}=-7(1)\times 10^{-2}$ $\frac{kJ}{m^3}$ for $Ni_{82}Fe_{18}$.  The sign is determined by the  magnetostrictive constant. The energetically favorable state for magnetization direction will therefore be along ($Co_{40}Fe_{40}B_{20}$) or perpendicular ($Ni$) to the direction of tensile strain. The strain induced effects are expected to be more than 50 times smaller in the devices made of $Ni_{82}Fe_{18}$.


\begin{figure}[h!]
\centering\includegraphics[width=8cm]{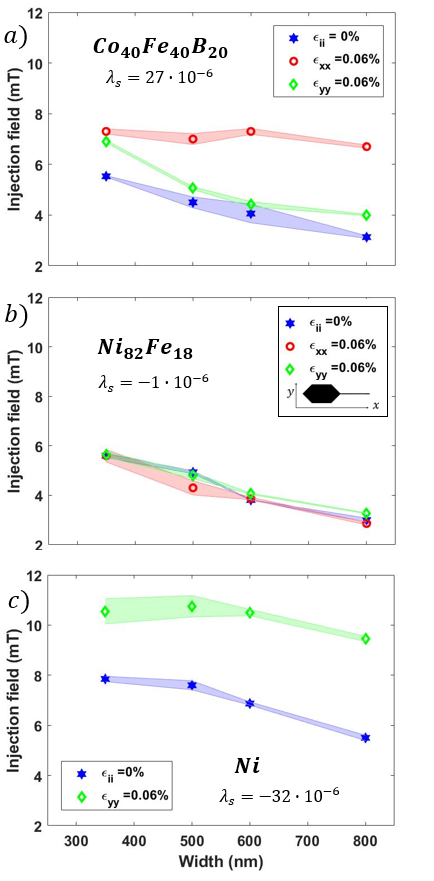}
\caption{\label{fig_exp_injection} Experimental results of the injection field $B_{inj}$. The experimental values are plotted as a function of the nominal width of the nanowire. Three different ferromagnetic materials (thickness 30 nm) have been measured: a) $Co_{40}Fe_{40}B_{20}$, b) $Ni_{82}Fe_{18}$ and c) $Ni$. For the data points in blue no strain is applied. Uniaxial strain $\epsilon_{ii}$ was applied in the x or in the y direction for the red and green curve, respectively as schematically shown in the inset. }
\end{figure}

In figure \ref{fig_exp_injection}, the injection field $B_{inj}$ is experimentally reported for three different materials. As expected, materials with strong magnetoelastic coupling (i.e. large $|\lambda_s|$) will show the largest strain effects as can be seen in figure \ref{fig_exp_injection} a) and c). This is why, for $Ni_{82}Fe_{18}$, with low magnetoelastic coupling, the three curves overlap within the error bars in figure \ref{fig_exp_injection} b). This also confirms that the observed changes in the injection field are caused by strain. 

A first observation is that in magnetostrictive materials strain is always increasing the injection field, thus imposing a lower limit for the sensor operation field. When an easy axis along $x$ is created, the pad coercivity grows, thus also increasing the injection field. This is the case of $\epsilon_{xx}$ for $Co_{40}Fe_{40}B_{20}$ and $\epsilon_{yy}$ for $Ni$. When instead the easy axis is oriented along the $y$ direction ($\epsilon_{yy}$ for $Co_{40}Fe_{40}B_{20}$ and $\epsilon_{xx}$ for $Ni$), the DW created at the mouth of the pad finds this position more energetically favorable than the wire. Therefore, a larger injection field is required. Due to small magnetic contrast in $Ni$, some experimental points in \ref{fig_exp_injection} c) are missing. 

A second observation concerns the wire width dependence of the injection field. Regardless of the material, when strain is not applied ($\epsilon_{xx}=\epsilon_{yy}=0 $), we find that the injection field decreases with the increasing width of the wire. As reported elsewhere\cite{im2009direct, bogart2009dependence}, the injection field  in soft magnetic wires is mainly determined by the shape of the cross section (width and thickness) and the monotonic behavior of $B_{inj}$ is the consequence of different sizes of DWs governed by different wire widths. Therefore, we expect that in the absence of strain the de-pinning field from the extremity of the pad is mainly due to shape anisotropy \cite{backes2007transverse}. However, when magnetoelastic anisotropy energy is introduced in the system, we observe deviation from this dependence. When $K_{ME}$ favors a spin orientation along the  $x$ direction, the injection field is determined by the coercivity of the pad and the dependence of the injection field on the wire width is negligible for thin wires (figure \ref{fig_exp_injection} a), red circles and \ref{fig_exp_injection} c),  green diamonds). This is because $B_c$ in the pad is large enough to inject and propagate the DW through the nanowire. In all the other measured cases, instead, the DW stays pinned at the edge of the pad, and a net dependence of $B_{inj}$ on the wire width is observable.  Interestingly, when the strain-induced easy axis of magnetization is perpendicular to the wire, the dependence of $B_{inj}$ on the wire width is stronger (figure \ref{fig_exp_injection} a), green diamonds). 


An explanation for this can be found in the different competing contributions to the system energy. To describe the impact of strain on the magnetization orientation and injection field, we use the free energy of the system $F_{tot}$ that is a measure for the angular dependence of the magnetic hardness. In a system with no net crystalline anisotropy, the free energy  is given by\cite{weiler2009voltage,brandlmaier2008situ}

\begin{equation} \label{eq_free_ener}
F_{tot}=F_{zeeman}+F_{demag}+F_{magel}.
\end{equation}

$F_{zeeman}$ describes the influence of the external magnetic field, and the demagnetization term $F_{demag}$ depends on the shape of the device. The last term describes the influence of the lattice strain to the magnetic anisotropy $F_{magel}=K_{ME}sin^2(\phi)$ according to equation \ref{eq_K_ME}, where $\phi$ is the angle between the magnetization and the easy axis. Minima in the expression of $F_{tot}$ correspond to magnetic easy directions. 

Let us now compare the case of $F_{magel}=0$ (no strain) and $K_{ME}^{y}\neq 0$ with the easy axis along y. This strain-induced uniaxial anisotropy tends to favor a spin configuration with wider DW in the nucleation pad, where a large part of magnetization is pointing along y. This is observed both in the MOKE images and in the simulated spin structure. What determines the injection field\cite{bogart2009dependence}, is the energy difference between a DW sitting at the extremity of the pad and inside the wire $\Delta E^{DW}=E^{DW}_{wire}-E^{DW}_{pad}$. Since in the wire a narrow DW is preferred\cite{mcmichael1997head}, the energy barrier $\Delta E^{DW}$ will be larger if $K_{ME}^{y}\neq 0$ . Therefore, a larger external applied field is required to inject the DW wall, as we experimentally observe.

To get a better understanding of the effect of strain on the injection field we performed micromagnetic simulations using the GPU-based $Mumax^3$ framework \cite{vansteenkiste2014design}. The material parameter values measured for our sample were considered, namely $\lambda_s=2.7\times 10^{-5}$ and $M_s=1.0\times 10^6 $ $A/m$  for saturation magnetostriction and magnetization, respectively, whereas for the exchange and elastic constants typical values for $Co_{40}Fe_{40}B_{20}$ reported in the literature were used  \cite{peng2016fast}: $A_{ex}=1.5\times10^{-11}$ $J/m$,  $c_{11}=2.8 \times10^{11}$ $N/m^2$, $c_{12}=1.4\times 10^{11}$ $N/m^2$ and $c_{44}=0.7\times 10^{11}$ $N/m^2$.


\begin{figure*}
\centering\includegraphics[width=15cm]{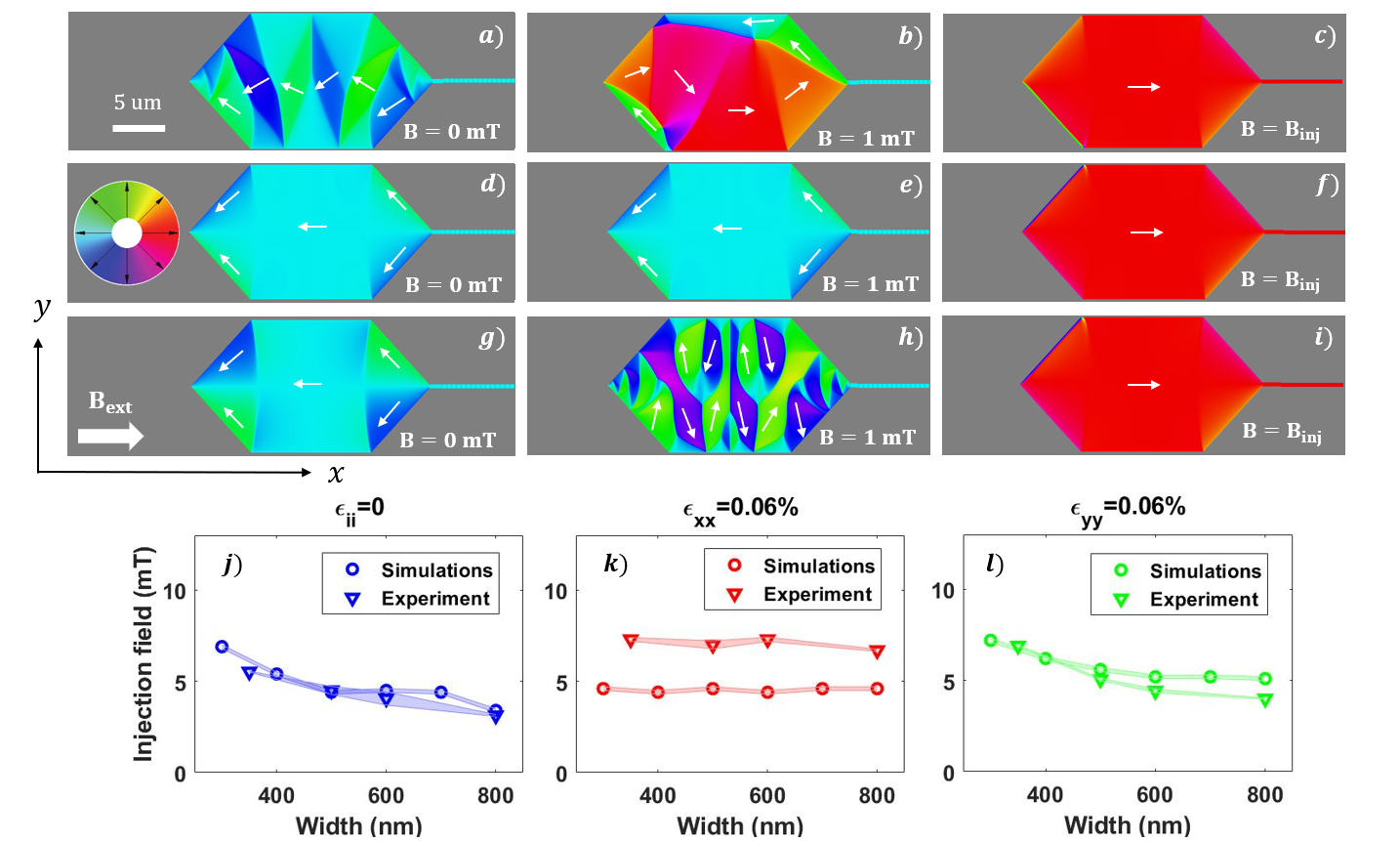}
\caption{\label{fig_simulations} 
 Snapshots of the simulations performed with the $Mumax^3$ framework.  We used material parameters of amorphous  $Co_{40}Fe_{40}B_{20}$. The field is applied along the $x$ direction and progressively increased (from left to right) until the DW nucleated in the pad is injected into the nanowire (300 nm wide). a) - c) strain is not included, d) - f) uniform tensile strain along the wire is applied,  $\epsilon_{xx}=0.06\%$, and g) - i) uniform tensile strain is applied perpendicular to the wire, $\epsilon_{yy}=0.06\%$.The order of the images follows figure \ref{fig_switching}, and a comparison confirms the same switching mechanism observed in the experiments. j) - l) comparison between the calculated and measured $B_{inj}$ for the indicated strain configurations.
}
\end{figure*}

In the simulations the dimensions of the pad are the same than in the physical system (figure \ref{fig_structure} c)), whereas the nanowire is shortened to $2.5$ $ \mu m$. The computational region is divided into $5\times5\times15$ $nm^3$ cells. The mechanical stress is modeled by adding a magnetoelastic field contribution to the effective field \cite{yanes2019skyrmion,hubert2008magnetic}. The system is initialized with uniform magnetization pointing in the $-x$ direction and the equilibrium state is calculated for a series of increasing applied fields in steps of $0.2$ $mT$. The results of the simulations are shown in figure \ref{fig_simulations}. Some snapshots of the magnetization are presented to be compared with the Kerr microscope images in figure \ref{fig_switching}. As noticeable, the main features observed at $B=1$ $mT$ are reproduced by the simulations, namely the double vortex pattern in the absence of strain (figures \ref{fig_switching} b) and \ref{fig_simulations} b)), the quasi-uniform configuration for $\epsilon_{xx}=0.06\%$ (figures \ref{fig_switching} e) and \ref{fig_simulations} e)) and the multidomain state favoring the magnetization pointing along y for $\epsilon_{yy}=0.06 \%$ (figures \ref{fig_switching} h) and \ref{fig_simulations} h)). The computed injection fields as a function of the wire width are plotted in the lower part of figure \ref{fig_simulations} together with the experimental ones, showing good quantitative agreement except for the case where the tensile strain is applied along x, where the computed values are significantly below the experimental ones. Both the decrease in the injection field when increasing the wire width for the cases of no strain (figure \ref{fig_simulations} j)) and $\epsilon_{yy}=0.06 \%$ (figure \ref{fig_simulations} l)) and the negligible dependence for $\epsilon_{xx}=0.06 \%$ are well captured by the simulations, supporting our interpretation of the experimental results. 

To summarize, an overview of the effects of the strain on the injection field can be observed in figure \ref{fig_exp_injection} a). We report that for a positive magnetostrictive material such as $Co_{40}Fe_{40}B_{20}$, $0.06\%$ uniaxial strain increases the injection field. In the case when the easy axis of magnetization is aligned along the wire, the injection field coincides with $B_c$ of the pad and the wire width dependence is low (figure \ref{fig_exp_injection} a), red circles). 


\subsection{Effects of growth induced anisotropy}

Up to now we have considered isotropic and magnetically soft ferromagnetic materials. In this case, a strain induced uniaxial anisotropy with constant of  $K_{ME}\simeq3-5$ $ \frac{kJ}{m^3}$ was the only anisotropy energy contribution in the full film material. To further investigate the mechanism and the limits for DW nucleation and injection, we structured our devices using thermally annealed $Co_{40}Fe_{40}B_{20}$. This material preparation induces a preferential orientation (easy axis, EA) for the magnetization, according to the direction of the applied magnetic field during annealing. The uniaxial anisotropy $K_{film}=1.54(2)$ $\frac{kJ}{m^3}$ has been measured experimentally from the full film hysteresis loops. To take this contribution into account, an additional term to the free energy is added, and equation \ref{eq_free_ener} becomes

\begin{equation} \label{eq_free_energy_andfilm}
F_{tot}=F_{zeeman}+F_{demag}+F_{magel}+F_{film},
\end{equation}

where $F_{film}=K_{film}sin^2(\phi)$ is the free energy term of the crystalline magnetic anisotropy, which may compete with strain-induced anisotropy and alter the magnetization orientation effects.


\begin{figure}[h!]
\centering\includegraphics[width=10cm]{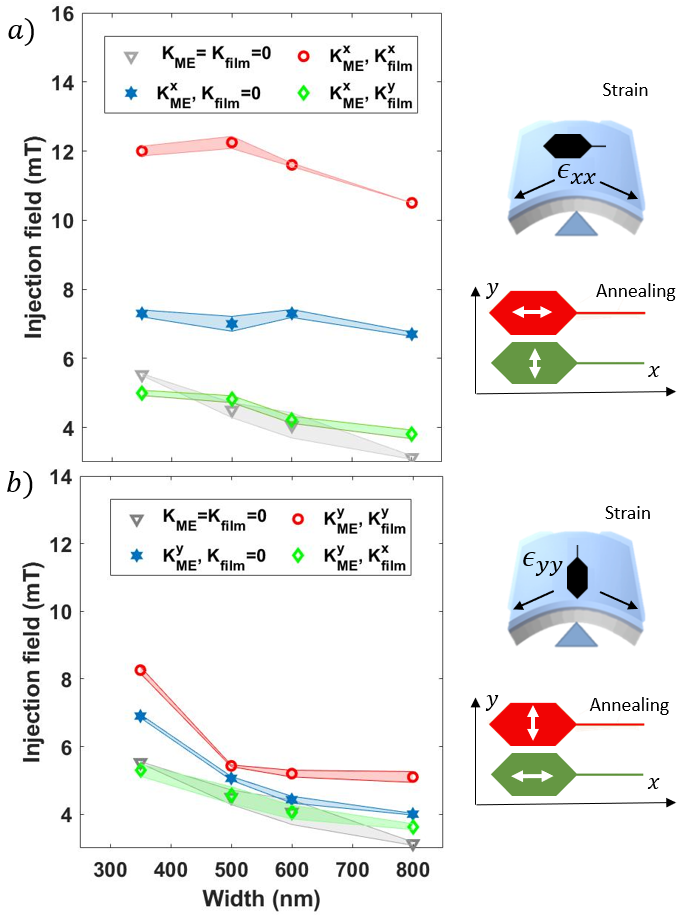}
\caption{\label{fig_anisotropy_strain} Experimental injection field for $Co_{40}Fe_{40}B_{20}$. Here samples with and without annealing are compared. a) $\epsilon_{xx}=0.06\%$ in plane strain is applied and b) $\epsilon_{yy}=0.06\%$. $K_{ME}$ and $K_{film}$ are the uniaxial magnetic anisotropy constants due to strain and annealing, respectively. The apexes indicate the in plane direction of the uniaxial anisotropy. Grey triangles are, for comparison, the reference case where no strain is applied. The scheme on the right shows the direction of the magnetic field applied during annealing (white arrows) relative to the tensile strain (black arrows).  }
\end{figure}

In light of this, the experiment described in the previous section has been repeated for annealed $Co_{40}Fe_{40}B_{20}$ samples. Sizable effects are found when $K_{film}$ and the magnetoelastic anisotropy $K_{ME}$ are superimposed. In figure \ref{fig_anisotropy_strain}, the results for this experiment are shown, and the annealed films are compared with the as-deposited samples. One can observe how, when the directions of $K_{ME}$ and $K_{film}$ are parallel, the effects of strain are enhanced (larger injection field for red points in figure \ref{fig_anisotropy_strain}). On the other hand, strain effects are attenuated or compensated if $K_{ME}$ and $K_{film}$ have perpendicular easy axis direction (green points in figure \ref{fig_anisotropy_strain}). When two different uniaxial anisotropy contributions point in the same direction, the resulting anisotropy of the film is still uniaxial, but now has an equivalent anisotropy constant  $K_{eq}\propto K_{ME}+K_{film}$. This is experimentally confirmed by an increase in the anisotropy field $B_{k}$, measured along the hard axis of magnetization. Again, we can distinguish two situations. When $K_{ME}^{x}$ and $K_{film}^{x}$ are oriented along $x$, the resulting EA is along the wire (figure \ref{fig_anisotropy_strain} a), red circles). This increases the coercive field of the pad and consequently the injection field also grows. When $K_{ME}^{y}$ and $K_{film}^{y}$ are oriented with EA along $y$, the resulting anisotropy favors energetically the DW to be positioned at the extremity of the pad. Consequently $\Delta E^{DW}$ and $B_{inj}$ are larger (figure \ref{fig_anisotropy_strain} b), red circles). 

The  nontrivial case is the situation when the two contributions of anisotropy, $K_{ME}$ and $K_{fim}$,  are perpendicular to each other. Experimentally, the values of the injection field are reduced and are close to the non-strained sample (gray triangles in figure \ref{fig_anisotropy_strain}). This result is important since it shows that strain effects on the device can be attenuated by material preparation. This outcome might seem unexpected, due to the difference in strength of the two contributions $K_{ME}=3.6$ $ \frac{kJ}{m^3}$ and  $K_{film}= 1.54$ $ \frac{kJ}{m^3}$. However, one should keep in mind that the idea of an "effective" uniaxial magnetic anisotropy  $K_{eq}$ is not applicable, unless the anisotropy are oriented along identical directions.

\begin{figure}[ht]
\centering\includegraphics[width=14cm]{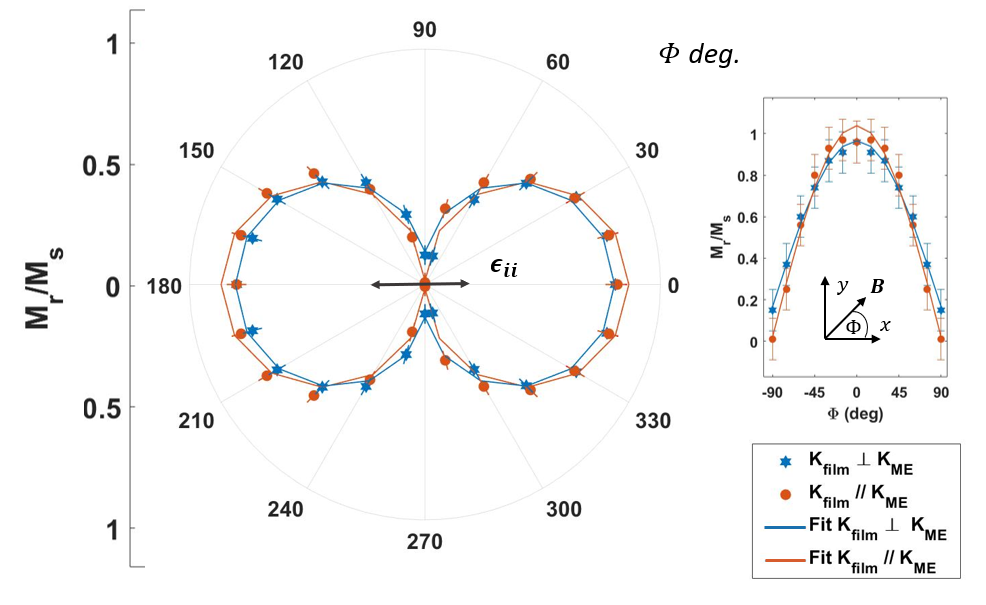}
\caption{\label{fig_perp_anis_film}  Angular dependence of the normalized remanent magnetization of full film $Co_{40}Fe_{40}B_{20}$. The material was strained along $\Phi=$0$^\circ$ and measured with MOKE contrast. In orange (circles) and blue (stars) the magnetic field during annealing was applied along $\Phi=$0$^\circ$ and 90$^\circ$ respectively. The experimental values were fitted using equation \ref{eq_perp} of $M_r$, and are here showed with lines. The offset of $M_r$ along the hard axis ($\Phi=$90$^\circ$) shows the overlap of two perpendicular uniaxial anisotropy directions (biaxial anisotropy), where the magnetoelastic contribution is dominant $K_{film}<K_{ME}$. In the inset, the data are displayed in a 2D plot.}
\end{figure}

To understand these results, a characterization  of the full film material has been done in the presence of strain and annealing-induced anisotropy with MOKE hysteresis loops. The most general case considers a magnetic energy described by two perpendicular uniaxial magnetic anisotropies axes. We have measured the angular  dependence of the  normalized remanent magnetization  $M_r/M_s$ as function of the angle $\Phi$  between the external magnetic field and the easy axis of magnetization. In figure \ref{fig_perp_anis_film} the cases $K_{ME}^{x}$ , $K_{film}^{x}$ and $K_{ME}^{x}$, $K_{film}^{y}$ are compared. In both cases, the strain is defining the dominant easy axis, since $K_{ME}>K_{film}$. However, in figure \ref{fig_perp_anis_film} we observe differences in the angular plots of $M_r$ in the vicinity of the hard axis ($\Phi=$90$^\circ$). The precise determination
of the magnetic easy axis is carried out by fitting the
normalized magnetic hysteresis $M_r/M_s$ curves as a function of $\Phi$. The projection of the in-plane magnetization vector to the plane of incidence of light during our MOKE measurements is a cosine-like function (as is evident from the inset in figure \ref{fig_perp_anis_film}), therefore, the following fitting function is chosen \cite{mollick2018strong}:

\begin{equation} \label{eq_perp}
\frac{M_r}{M_s}=\frac{M_r^{max}}{M_s} |cos(\Phi)|+\frac{M_r^{off}}{M_s}
\end{equation}

where $\frac{M_r^{max}}{M_s}$ is the maximum normalized magnetic remanence, and $\frac{M_r^{off}}{M_s}$ is the offset in magnetic remanence $M_r$. The strength of the uniaxial magnetic anisotropy is the amplitude of the fitting parameter, and the offset originates from the isotropic contribution of the film.


The fitting confirms that the easy axis is along $\Phi=0^\circ$ (strain direction). However, the strength of the dominant magnetic easy axis is strongly reduced when  $K_{ME} \perp K_{film}$. More quantitatively, if we compare the two cases $K_{ME} \parallel K_{film}$ and $K_{ME} \perp K_{film}$, the offset $\frac{M_r^{off}}{M_s}$ increases from $0.02$ to $0.21$ while the maximum remanence decreases from $ 1.00$ to $0.75$, respectively. This means the annealing induced anisotropy can contribute to the effective magnetic anisotropy of the system. The pure uniaxial anisotropy is modified by the presence of an isotropic part. 





The observed results and consideration on the full film material properties can explain the experimentally determined injection field. In figure \ref{fig_anisotropy_strain}, we showed how the effects of strain can be compensated by annealing $Co_{40}Fe_{40}B_{20}$ with a magnetic field perpendicular to the strain direction. In a material with  larger $\frac{M_r^{off}}{M_s}$ the coherent rotation of the magnetization requires less energy: the anisotropy field decreases and the coercive field is reduced at the same time. Therefore, the film becomes magnetically softer, and the magnetic properties of the blue points in figure \ref{fig_perp_anis_film} are closer to the ones of the unstrained as-deposited state, shown in figure \ref{fig_loops} a). This supports the findings displayed in figure \ref{fig_anisotropy_strain}, where the green ($K_{ME} \perp K_{film}$) and gray (no anisotropy) data points overlap.

The material softness directly influences not only the creation and nucleation of the DW in the nucleation pad, but also the injection field, due to the fact that the energy difference between the DW at the mouth of the pad and in the nanowire is affected by the film properties. We can indeed say that the DW energy in the thin long wire is unchanged by the presence of an annealing-induced anisotropy, because here the strong shape anisotropy\cite{kumar2006competition} is in the order of $\simeq 10^{4} $ $J/m^3$. On the other hand, in the nucleation pad, if the strength of the uniaxial magnetic anisotropy is reduced, the magnetization is free to rotate in the field direction similarly to the unstrained as-deposited state.



\section{Conclusions}

In conclusion, we have investigated the effects of  mechanical strain on the injection of a DW into a nanowire using MOKE microscopy and $Mumax^3$ micromagnetic simulations. We have measured in-plane magnetized $Co_{40}Fe_{40}B_{20}$ , $Ni$ and $Ni_{82}Fe_{18}$ films structured by optical lithography. We find that the effects of the strain are proportional to the magnetoelastic coupling of the material, quantified by the saturation magnetostriction $\lambda_s$. We report that uniaxial strain, regardless of the direction, induces a uniaxial magnetic anisotropy in the material and increases the injection field. The changes are up to $30\%$ in materials with $|\lambda_s|\simeq 30$ $p.p.m.$, while are negligible in low magnetostrictive $Ni_{82}Fe_{18}$. The experimental results  show how the mechanism of DW injection depends primarily on the creation of the DW in the pad. Further measurements involved  different material preparation introducing a thermal annealing step with a magnetic field for $Co_{40}Fe_{40}B_{20}$. With this, another uniaxial magnetic anisotropy is introduced, and contributes together with strain effects to the $B_{inj}$. We find that the strain induced uniaxial anisotropy, $K_{ME}$,  can be enhanced or weakened using annealing induced uniaxial anisotropy, $K_{film}$, respectively, parallel or perpendicular to the strain contribution. If the easy axis of magnetization is along the wire direction, the coercive field of the nucleation pad increases significantly. In the same way, low anisotropy will facilitate the magnetization to switch, and creates a DW at low fields ($\simeq1mT$) at the extremity of the pad.  The DW stays pinned at the wire entrance until the energy difference $\Delta E^{DW}=E^{DW}_{wire}-E^{DW}_{pad}$ is comparable with the Zeeman energy (higher external fields). This energy barrier is the combination of different anisotropies, and can be tailored by device design and material preparation. The validity of these results is verified by micromagnetic simulations, that  can help to identify the lower limit for reliable DW injection. The optimization and development of magnetic sensors and devices based on domain walls needs to consider, therefore, the effects of strain and material preparation. Our results of the DW injection mechanism show that a magnetostrictive free behavior of the DW based device, can even be reached in systems with finite magnetostriction. A careful material preparation, can reduce the effective anisotropy caused by strain in the magnetic layer thus keeping the DW injection field low in these devices. This provides, therefore, a way to improve robustness of these type of magnetic sensors against strain disturbances.



\begin{acknowledgments}

The authors would like to acknowledge Dr.   R. Lehndorff for the fruitful discussions.   This project has received funding from the European Union’s Horizon 2020 research and innovation program  under  the  Marie  Skłodowska-Curie  grant  agreement  No  860060  “Magnetism  and  the effect of Electric Field” (MagnEFi) and the Austrian Research Promotion Agency (FFG). The authors also acknowledge support by the chip production facilities of Sensitec GmbH (Mainz, DE), where part of this work was carried out and the Max-Planck Graduate Centre with Johannes Gutenberg University.
\end{acknowledgments}

\section*{Data Sharing Policy }
The data that support the findings of this study are available from the corresponding author upon reasonable request. This article has been submitted to Journal of Applied Physics. After it is published, it will be found on the AIP Publishing website .





\nocite{*}
\bibliography{aipsamp}

\providecommand{\noopsort}[1]{}\providecommand{\singleletter}[1]{#1}%
\begin{thebibliography}{41}%
\makeatletter
\providecommand \@ifxundefined [1]{%
 \@ifx{#1\undefined}
}%
\providecommand \@ifnum [1]{%
 \ifnum #1\expandafter \@firstoftwo
 \else \expandafter \@secondoftwo
 \fi
}%
\providecommand \@ifx [1]{%
 \ifx #1\expandafter \@firstoftwo
 \else \expandafter \@secondoftwo
 \fi
}%
\providecommand \natexlab [1]{#1}%
\providecommand \enquote  [1]{``#1''}%
\providecommand \bibnamefont  [1]{#1}%
\providecommand \bibfnamefont [1]{#1}%
\providecommand \citenamefont [1]{#1}%
\providecommand \href@noop [0]{\@secondoftwo}%
\providecommand \href [0]{\begingroup \@sanitize@url \@href}%
\providecommand \@href[1]{\@@startlink{#1}\@@href}%
\providecommand \@@href[1]{\endgroup#1\@@endlink}%
\providecommand \@sanitize@url [0]{\catcode `\\12\catcode `\$12\catcode
  `\&12\catcode `\#12\catcode `\^12\catcode `\_12\catcode `\%12\relax}%
\providecommand \@@startlink[1]{}%
\providecommand \@@endlink[0]{}%
\providecommand \url  [0]{\begingroup\@sanitize@url \@url }%
\providecommand \@url [1]{\endgroup\@href {#1}{\urlprefix }}%
\providecommand \urlprefix  [0]{URL }%
\providecommand \Eprint [0]{\href }%
\providecommand \doibase [0]{http://dx.doi.org/}%
\providecommand \selectlanguage [0]{\@gobble}%
\providecommand \bibinfo  [0]{\@secondoftwo}%
\providecommand \bibfield  [0]{\@secondoftwo}%
\providecommand \translation [1]{[#1]}%
\providecommand \BibitemOpen [0]{}%
\providecommand \bibitemStop [0]{}%
\providecommand \bibitemNoStop [0]{.\EOS\space}%
\providecommand \EOS [0]{\spacefactor3000\relax}%
\providecommand \BibitemShut  [1]{\csname bibitem#1\endcsname}%
\let\auto@bib@innerbib\@empty
\bibitem [{\citenamefont {Kl{\"a}ui}(2008)}]{klaui2008head}%
  \BibitemOpen
  \bibfield  {author} {\bibinfo {author} {\bibfnamefont {M.}~\bibnamefont
  {Kl{\"a}ui}},\ }\bibfield  {title} {\enquote {\bibinfo {title} {Head-to-head
  domain walls in magnetic nanostructures},}\ }\href@noop {} {\bibfield
  {journal} {\bibinfo  {journal} {Journal of Physics: Condensed matter}\
  }\textbf {\bibinfo {volume} {20}},\ \bibinfo {pages} {313001} (\bibinfo
  {year} {2008})}\BibitemShut {NoStop}%
\bibitem [{\citenamefont {Spain}(1966)}]{1065929}%
  \BibitemOpen
  \bibfield  {author} {\bibinfo {author} {\bibfnamefont {R.}~\bibnamefont
  {Spain}},\ }\bibfield  {title} {\enquote {\bibinfo {title} {Domain tip
  propagation logic},}\ }\href {\doibase 10.1109/TMAG.1966.1065929} {\bibfield
  {journal} {\bibinfo  {journal} {IEEE Transactions on Magnetics}\ }\textbf
  {\bibinfo {volume} {2}},\ \bibinfo {pages} {347--351} (\bibinfo {year}
  {1966})}\BibitemShut {NoStop}%
\bibitem [{\citenamefont {Parkin}, \citenamefont {Hayashi},\ and\ \citenamefont
  {Thomas}(2008)}]{parkin2008magnetic}%
  \BibitemOpen
  \bibfield  {author} {\bibinfo {author} {\bibfnamefont {S.~S.}\ \bibnamefont
  {Parkin}}, \bibinfo {author} {\bibfnamefont {M.}~\bibnamefont {Hayashi}}, \
  and\ \bibinfo {author} {\bibfnamefont {L.}~\bibnamefont {Thomas}},\
  }\bibfield  {title} {\enquote {\bibinfo {title} {Magnetic domain-wall
  racetrack memory},}\ }\href@noop {} {\bibfield  {journal} {\bibinfo
  {journal} {Science}\ }\textbf {\bibinfo {volume} {320}},\ \bibinfo {pages}
  {190--194} (\bibinfo {year} {2008})}\BibitemShut {NoStop}%
\bibitem [{\citenamefont {Jogschies}\ \emph {et~al.}(2015)\citenamefont
  {Jogschies}, \citenamefont {Klaas}, \citenamefont {Kruppe}, \citenamefont
  {Rittinger}, \citenamefont {Taptimthong}, \citenamefont {Wienecke},
  \citenamefont {Rissing},\ and\ \citenamefont {Wurz}}]{jogschies2015recent}%
  \BibitemOpen
  \bibfield  {author} {\bibinfo {author} {\bibfnamefont {L.}~\bibnamefont
  {Jogschies}}, \bibinfo {author} {\bibfnamefont {D.}~\bibnamefont {Klaas}},
  \bibinfo {author} {\bibfnamefont {R.}~\bibnamefont {Kruppe}}, \bibinfo
  {author} {\bibfnamefont {J.}~\bibnamefont {Rittinger}}, \bibinfo {author}
  {\bibfnamefont {P.}~\bibnamefont {Taptimthong}}, \bibinfo {author}
  {\bibfnamefont {A.}~\bibnamefont {Wienecke}}, \bibinfo {author}
  {\bibfnamefont {L.}~\bibnamefont {Rissing}}, \ and\ \bibinfo {author}
  {\bibfnamefont {M.~C.}\ \bibnamefont {Wurz}},\ }\bibfield  {title} {\enquote
  {\bibinfo {title} {Recent developments of magnetoresistive sensors for
  industrial applications},}\ }\href@noop {} {\bibfield  {journal} {\bibinfo
  {journal} {Sensors}\ }\textbf {\bibinfo {volume} {15}},\ \bibinfo {pages}
  {28665--28689} (\bibinfo {year} {2015})}\BibitemShut {NoStop}%
\bibitem [{\citenamefont {Diegel}, \citenamefont {Mattheis},\ and\
  \citenamefont {Halder}(2007)}]{diegel2007multiturn}%
  \BibitemOpen
  \bibfield  {author} {\bibinfo {author} {\bibfnamefont {M.}~\bibnamefont
  {Diegel}}, \bibinfo {author} {\bibfnamefont {R.}~\bibnamefont {Mattheis}}, \
  and\ \bibinfo {author} {\bibfnamefont {E.}~\bibnamefont {Halder}},\
  }\bibfield  {title} {\enquote {\bibinfo {title} {Multiturn counter using
  movement and storage of 180° magnetic domain walls},}\ }\href@noop {}
  {\bibfield  {journal} {\bibinfo  {journal} {Sensor Letters}\ }\textbf
  {\bibinfo {volume} {5}},\ \bibinfo {pages} {118--122} (\bibinfo {year}
  {2007})}\BibitemShut {NoStop}%
\bibitem [{\citenamefont {Mattheis}\ \emph {et~al.}(2012)\citenamefont
  {Mattheis}, \citenamefont {Glathe}, \citenamefont {Diegel},\ and\
  \citenamefont {H{\"u}bner}}]{mattheis2012concepts}%
  \BibitemOpen
  \bibfield  {author} {\bibinfo {author} {\bibfnamefont {R.}~\bibnamefont
  {Mattheis}}, \bibinfo {author} {\bibfnamefont {S.}~\bibnamefont {Glathe}},
  \bibinfo {author} {\bibfnamefont {M.}~\bibnamefont {Diegel}}, \ and\ \bibinfo
  {author} {\bibfnamefont {U.}~\bibnamefont {H{\"u}bner}},\ }\bibfield  {title}
  {\enquote {\bibinfo {title} {Concepts and steps for the realization of a new
  domain wall based giant magnetoresistance nanowire device: From the available
  24 multiturn counter to a 212 turn counter},}\ }\href@noop {} {\bibfield
  {journal} {\bibinfo  {journal} {Journal of Applied Physics}\ }\textbf
  {\bibinfo {volume} {111}},\ \bibinfo {pages} {113920} (\bibinfo {year}
  {2012})}\BibitemShut {NoStop}%
\bibitem [{\citenamefont {Borie}\ \emph
  {et~al.}(2017{\natexlab{a}})\citenamefont {Borie}, \citenamefont {Wahrhusen},
  \citenamefont {Grimm},\ and\ \citenamefont
  {Kl{\"a}ui}}]{borie2017geometrically}%
  \BibitemOpen
  \bibfield  {author} {\bibinfo {author} {\bibfnamefont {B.}~\bibnamefont
  {Borie}}, \bibinfo {author} {\bibfnamefont {J.}~\bibnamefont {Wahrhusen}},
  \bibinfo {author} {\bibfnamefont {H.}~\bibnamefont {Grimm}}, \ and\ \bibinfo
  {author} {\bibfnamefont {M.}~\bibnamefont {Kl{\"a}ui}},\ }\bibfield  {title}
  {\enquote {\bibinfo {title} {Geometrically enhanced closed-loop multi-turn
  sensor devices that enable reliable magnetic domain wall motion},}\
  }\href@noop {} {\bibfield  {journal} {\bibinfo  {journal} {Applied Physics
  Letters}\ }\textbf {\bibinfo {volume} {111}},\ \bibinfo {pages} {242402}
  (\bibinfo {year} {2017}{\natexlab{a}})}\BibitemShut {NoStop}%
\bibitem [{\citenamefont {Borie}\ \emph
  {et~al.}(2017{\natexlab{b}})\citenamefont {Borie}, \citenamefont
  {Kehlberger}, \citenamefont {Wahrhusen}, \citenamefont {Grimm},\ and\
  \citenamefont {Kl{\"a}ui}}]{borie2017geometrical}%
  \BibitemOpen
  \bibfield  {author} {\bibinfo {author} {\bibfnamefont {B.}~\bibnamefont
  {Borie}}, \bibinfo {author} {\bibfnamefont {A.}~\bibnamefont {Kehlberger}},
  \bibinfo {author} {\bibfnamefont {J.}~\bibnamefont {Wahrhusen}}, \bibinfo
  {author} {\bibfnamefont {H.}~\bibnamefont {Grimm}}, \ and\ \bibinfo {author}
  {\bibfnamefont {M.}~\bibnamefont {Kl{\"a}ui}},\ }\bibfield  {title} {\enquote
  {\bibinfo {title} {Geometrical dependence of domain-wall propagation and
  nucleation fields in magnetic-domain-wall sensors},}\ }\href@noop {}
  {\bibfield  {journal} {\bibinfo  {journal} {Physical Review Applied}\
  }\textbf {\bibinfo {volume} {8}},\ \bibinfo {pages} {024017} (\bibinfo {year}
  {2017}{\natexlab{b}})}\BibitemShut {NoStop}%
\bibitem [{\citenamefont {Martinez}\ \emph {et~al.}(2007)\citenamefont
  {Martinez}, \citenamefont {Lopez-Diaz}, \citenamefont {Alejos}, \citenamefont
  {Torres},\ and\ \citenamefont {Tristan}}]{martinez2007thermal}%
  \BibitemOpen
  \bibfield  {author} {\bibinfo {author} {\bibfnamefont {E.}~\bibnamefont
  {Martinez}}, \bibinfo {author} {\bibfnamefont {L.}~\bibnamefont
  {Lopez-Diaz}}, \bibinfo {author} {\bibfnamefont {O.}~\bibnamefont {Alejos}},
  \bibinfo {author} {\bibfnamefont {L.}~\bibnamefont {Torres}}, \ and\ \bibinfo
  {author} {\bibfnamefont {C.}~\bibnamefont {Tristan}},\ }\bibfield  {title}
  {\enquote {\bibinfo {title} {Thermal effects on domain wall depinning from a
  single notch},}\ }\href@noop {} {\bibfield  {journal} {\bibinfo  {journal}
  {Physical Review Letters}\ }\textbf {\bibinfo {volume} {98}},\ \bibinfo
  {pages} {267202} (\bibinfo {year} {2007})}\BibitemShut {NoStop}%
\bibitem [{\citenamefont {Martinez}\ \emph {et~al.}(2009)\citenamefont
  {Martinez}, \citenamefont {Lopez-Diaz}, \citenamefont {Alejos}, \citenamefont
  {Torres},\ and\ \citenamefont {Carpentieri}}]{martinez2009domain}%
  \BibitemOpen
  \bibfield  {author} {\bibinfo {author} {\bibfnamefont {E.}~\bibnamefont
  {Martinez}}, \bibinfo {author} {\bibfnamefont {L.}~\bibnamefont
  {Lopez-Diaz}}, \bibinfo {author} {\bibfnamefont {O.}~\bibnamefont {Alejos}},
  \bibinfo {author} {\bibfnamefont {L.}~\bibnamefont {Torres}}, \ and\ \bibinfo
  {author} {\bibfnamefont {M.}~\bibnamefont {Carpentieri}},\ }\bibfield
  {title} {\enquote {\bibinfo {title} {Domain-wall dynamics driven by short
  pulses along thin ferromagnetic strips: Micromagnetic simulations and
  analytical description},}\ }\href@noop {} {\bibfield  {journal} {\bibinfo
  {journal} {Physical Review B}\ }\textbf {\bibinfo {volume} {79}},\ \bibinfo
  {pages} {094430} (\bibinfo {year} {2009})}\BibitemShut {NoStop}%
\bibitem [{\citenamefont {Garcia-Sanchez}\ \emph {et~al.}(2011)\citenamefont
  {Garcia-Sanchez}, \citenamefont {Kakay}, \citenamefont {Hertel},\ and\
  \citenamefont {Asselin}}]{garcia2011depinning}%
  \BibitemOpen
  \bibfield  {author} {\bibinfo {author} {\bibfnamefont {F.}~\bibnamefont
  {Garcia-Sanchez}}, \bibinfo {author} {\bibfnamefont {A.}~\bibnamefont
  {Kakay}}, \bibinfo {author} {\bibfnamefont {R.}~\bibnamefont {Hertel}}, \
  and\ \bibinfo {author} {\bibfnamefont {P.}~\bibnamefont {Asselin}},\
  }\bibfield  {title} {\enquote {\bibinfo {title} {Depinning of transverse
  domain walls from notches in magnetostatically coupled nanostrips},}\
  }\href@noop {} {\bibfield  {journal} {\bibinfo  {journal} {Applied physics
  express}\ }\textbf {\bibinfo {volume} {4}},\ \bibinfo {pages} {033001}
  (\bibinfo {year} {2011})}\BibitemShut {NoStop}%
\bibitem [{\citenamefont {Hoang}\ \emph {et~al.}(2020)\citenamefont {Hoang},
  \citenamefont {Cao}, \citenamefont {Nguyen},\ and\ \citenamefont
  {Dao}}]{hoang2020creation}%
  \BibitemOpen
  \bibfield  {author} {\bibinfo {author} {\bibfnamefont {D.-Q.}\ \bibnamefont
  {Hoang}}, \bibinfo {author} {\bibfnamefont {X.-H.}\ \bibnamefont {Cao}},
  \bibinfo {author} {\bibfnamefont {H.-T.}\ \bibnamefont {Nguyen}}, \ and\
  \bibinfo {author} {\bibfnamefont {V.-A.}\ \bibnamefont {Dao}},\ }\bibfield
  {title} {\enquote {\bibinfo {title} {Creation and propagation of a single
  magnetic domain wall in {2D} nanotraps with a square injection pad},}\
  }\href@noop {} {\bibfield  {journal} {\bibinfo  {journal} {Nanotechnology}\
  }\textbf {\bibinfo {volume} {32}},\ \bibinfo {pages} {095703} (\bibinfo
  {year} {2020})}\BibitemShut {NoStop}%
\bibitem [{\citenamefont {Van~Driel}\ \emph {et~al.}(2003)\citenamefont
  {Van~Driel}, \citenamefont {Janssen}, \citenamefont {Zhang}, \citenamefont
  {Yang},\ and\ \citenamefont {Ernst}}]{van2003packaging}%
  \BibitemOpen
  \bibfield  {author} {\bibinfo {author} {\bibfnamefont {W.}~\bibnamefont
  {Van~Driel}}, \bibinfo {author} {\bibfnamefont {J.}~\bibnamefont {Janssen}},
  \bibinfo {author} {\bibfnamefont {G.}~\bibnamefont {Zhang}}, \bibinfo
  {author} {\bibfnamefont {D.}~\bibnamefont {Yang}}, \ and\ \bibinfo {author}
  {\bibfnamefont {L.}~\bibnamefont {Ernst}},\ }\bibfield  {title} {\enquote
  {\bibinfo {title} {Packaging induced die stresses-effect of chip anisotropy
  and time-dependent behavior of a molding compound},}\ }\href@noop {}
  {\bibfield  {journal} {\bibinfo  {journal} {Journal of Electronic Packaging}\
  }\textbf {\bibinfo {volume} {125}},\ \bibinfo {pages} {520--526} (\bibinfo
  {year} {2003})}\BibitemShut {NoStop}%
\bibitem [{\citenamefont {Lee}(1955)}]{lee1955magnetostriction}%
  \BibitemOpen
  \bibfield  {author} {\bibinfo {author} {\bibfnamefont {E.~W.}\ \bibnamefont
  {Lee}},\ }\bibfield  {title} {\enquote {\bibinfo {title} {Magnetostriction
  and magnetomechanical effects},}\ }\href@noop {} {\bibfield  {journal}
  {\bibinfo  {journal} {Reports on Progress in Physics}\ }\textbf {\bibinfo
  {volume} {18}},\ \bibinfo {pages} {184} (\bibinfo {year} {1955})}\BibitemShut
  {NoStop}%
\bibitem [{\citenamefont {Finizio}\ \emph {et~al.}(2014)\citenamefont
  {Finizio}, \citenamefont {Foerster}, \citenamefont {Buzzi}, \citenamefont
  {Kr{\"u}ger}, \citenamefont {Jourdan}, \citenamefont {Vaz}, \citenamefont
  {Hockel}, \citenamefont {Miyawaki}, \citenamefont {Tkach}, \citenamefont
  {Valencia} \emph {et~al.}}]{finizio2014magnetic}%
  \BibitemOpen
  \bibfield  {author} {\bibinfo {author} {\bibfnamefont {S.}~\bibnamefont
  {Finizio}}, \bibinfo {author} {\bibfnamefont {M.}~\bibnamefont {Foerster}},
  \bibinfo {author} {\bibfnamefont {M.}~\bibnamefont {Buzzi}}, \bibinfo
  {author} {\bibfnamefont {B.}~\bibnamefont {Kr{\"u}ger}}, \bibinfo {author}
  {\bibfnamefont {M.}~\bibnamefont {Jourdan}}, \bibinfo {author} {\bibfnamefont
  {C.~A.}\ \bibnamefont {Vaz}}, \bibinfo {author} {\bibfnamefont
  {J.}~\bibnamefont {Hockel}}, \bibinfo {author} {\bibfnamefont
  {T.}~\bibnamefont {Miyawaki}}, \bibinfo {author} {\bibfnamefont
  {A.}~\bibnamefont {Tkach}}, \bibinfo {author} {\bibfnamefont
  {S.}~\bibnamefont {Valencia}},  \emph {et~al.},\ }\bibfield  {title}
  {\enquote {\bibinfo {title} {Magnetic anisotropy engineering in thin film
  {Ni} nanostructures by magnetoelastic coupling},}\ }\href@noop {} {\bibfield
  {journal} {\bibinfo  {journal} {Physical Review Applied}\ }\textbf {\bibinfo
  {volume} {1}},\ \bibinfo {pages} {021001} (\bibinfo {year}
  {2014})}\BibitemShut {NoStop}%
\bibitem [{\citenamefont {Lei}\ \emph {et~al.}(2013)\citenamefont {Lei},
  \citenamefont {Devolder}, \citenamefont {Agnus}, \citenamefont {Aubert},
  \citenamefont {Daniel}, \citenamefont {Kim}, \citenamefont {Zhao},
  \citenamefont {Trypiniotis}, \citenamefont {Cowburn}, \citenamefont
  {Chappert} \emph {et~al.}}]{lei2013strain}%
  \BibitemOpen
  \bibfield  {author} {\bibinfo {author} {\bibfnamefont {N.}~\bibnamefont
  {Lei}}, \bibinfo {author} {\bibfnamefont {T.}~\bibnamefont {Devolder}},
  \bibinfo {author} {\bibfnamefont {G.}~\bibnamefont {Agnus}}, \bibinfo
  {author} {\bibfnamefont {P.}~\bibnamefont {Aubert}}, \bibinfo {author}
  {\bibfnamefont {L.}~\bibnamefont {Daniel}}, \bibinfo {author} {\bibfnamefont
  {J.-V.}\ \bibnamefont {Kim}}, \bibinfo {author} {\bibfnamefont
  {W.}~\bibnamefont {Zhao}}, \bibinfo {author} {\bibfnamefont {T.}~\bibnamefont
  {Trypiniotis}}, \bibinfo {author} {\bibfnamefont {R.~P.}\ \bibnamefont
  {Cowburn}}, \bibinfo {author} {\bibfnamefont {C.}~\bibnamefont {Chappert}},
  \emph {et~al.},\ }\bibfield  {title} {\enquote {\bibinfo {title}
  {Strain-controlled magnetic domain wall propagation in hybrid
  piezoelectric/ferromagnetic structures},}\ }\href@noop {} {\bibfield
  {journal} {\bibinfo  {journal} {Nature Communications}\ }\textbf {\bibinfo
  {volume} {4}},\ \bibinfo {pages} {1--7} (\bibinfo {year} {2013})}\BibitemShut
  {NoStop}%
\bibitem [{\citenamefont {Cowburn}\ \emph {et~al.}(2002)\citenamefont
  {Cowburn}, \citenamefont {Allwood}, \citenamefont {Xiong},\ and\
  \citenamefont {Cooke}}]{cowburn2002domain}%
  \BibitemOpen
  \bibfield  {author} {\bibinfo {author} {\bibfnamefont {R.}~\bibnamefont
  {Cowburn}}, \bibinfo {author} {\bibfnamefont {D.}~\bibnamefont {Allwood}},
  \bibinfo {author} {\bibfnamefont {G.}~\bibnamefont {Xiong}}, \ and\ \bibinfo
  {author} {\bibfnamefont {M.}~\bibnamefont {Cooke}},\ }\bibfield  {title}
  {\enquote {\bibinfo {title} {Domain wall injection and propagation in planar
  permalloy nanowires},}\ }\href@noop {} {\bibfield  {journal} {\bibinfo
  {journal} {Journal of Applied Physics}\ }\textbf {\bibinfo {volume} {91}},\
  \bibinfo {pages} {6949--6951} (\bibinfo {year} {2002})}\BibitemShut {NoStop}%
\bibitem [{\citenamefont {Shigeto}, \citenamefont {Shinjo},\ and\ \citenamefont
  {Ono}(1999)}]{shigeto1999injection}%
  \BibitemOpen
  \bibfield  {author} {\bibinfo {author} {\bibfnamefont {K.}~\bibnamefont
  {Shigeto}}, \bibinfo {author} {\bibfnamefont {T.}~\bibnamefont {Shinjo}}, \
  and\ \bibinfo {author} {\bibfnamefont {T.}~\bibnamefont {Ono}},\ }\bibfield
  {title} {\enquote {\bibinfo {title} {Injection of a magnetic domain wall into
  a submicron magnetic wire},}\ }\href@noop {} {\bibfield  {journal} {\bibinfo
  {journal} {Applied Physics Letters}\ }\textbf {\bibinfo {volume} {75}},\
  \bibinfo {pages} {2815--2817} (\bibinfo {year} {1999})}\BibitemShut {NoStop}%
\bibitem [{\citenamefont {Zhou}\ \emph {et~al.}(2020)\citenamefont {Zhou},
  \citenamefont {Shi}, \citenamefont {Nian}, \citenamefont {Cui}, \citenamefont
  {Luo}, \citenamefont {Qiu}, \citenamefont {Yang}, \citenamefont {Zhu},\ and\
  \citenamefont {Yu}}]{zhou2020voltage}%
  \BibitemOpen
  \bibfield  {author} {\bibinfo {author} {\bibfnamefont {H.}~\bibnamefont
  {Zhou}}, \bibinfo {author} {\bibfnamefont {S.}~\bibnamefont {Shi}}, \bibinfo
  {author} {\bibfnamefont {D.}~\bibnamefont {Nian}}, \bibinfo {author}
  {\bibfnamefont {S.}~\bibnamefont {Cui}}, \bibinfo {author} {\bibfnamefont
  {J.}~\bibnamefont {Luo}}, \bibinfo {author} {\bibfnamefont {Y.}~\bibnamefont
  {Qiu}}, \bibinfo {author} {\bibfnamefont {H.}~\bibnamefont {Yang}}, \bibinfo
  {author} {\bibfnamefont {M.}~\bibnamefont {Zhu}}, \ and\ \bibinfo {author}
  {\bibfnamefont {G.}~\bibnamefont {Yu}},\ }\bibfield  {title} {\enquote
  {\bibinfo {title} {Voltage control of magnetic domain wall injection into
  strain-mediated multiferroic heterostructures},}\ }\href@noop {} {\bibfield
  {journal} {\bibinfo  {journal} {Nanoscale}\ }\textbf {\bibinfo {volume}
  {12}},\ \bibinfo {pages} {14479--14486} (\bibinfo {year} {2020})}\BibitemShut
  {NoStop}%
\bibitem [{\citenamefont {Choe}\ and\ \citenamefont
  {Megdal}(1999)}]{choe1999high}%
  \BibitemOpen
  \bibfield  {author} {\bibinfo {author} {\bibfnamefont {G.}~\bibnamefont
  {Choe}}\ and\ \bibinfo {author} {\bibfnamefont {B.}~\bibnamefont {Megdal}},\
  }\bibfield  {title} {\enquote {\bibinfo {title} {High precision
  magnetostriction measurement employing the {BH} looper bending method},}\
  }\href@noop {} {\bibfield  {journal} {\bibinfo  {journal} {IEEE Transactions
  on Magnetics}\ }\textbf {\bibinfo {volume} {35}},\ \bibinfo {pages}
  {3959--3961} (\bibinfo {year} {1999})}\BibitemShut {NoStop}%
\bibitem [{\citenamefont {Hill}\ \emph {et~al.}(2013)\citenamefont {Hill},
  \citenamefont {Hendren}, \citenamefont {Bowman}, \citenamefont {McGeehin},
  \citenamefont {Gubbins},\ and\ \citenamefont {Venugopal}}]{hill2013whole}%
  \BibitemOpen
  \bibfield  {author} {\bibinfo {author} {\bibfnamefont {C.}~\bibnamefont
  {Hill}}, \bibinfo {author} {\bibfnamefont {W.}~\bibnamefont {Hendren}},
  \bibinfo {author} {\bibfnamefont {R.}~\bibnamefont {Bowman}}, \bibinfo
  {author} {\bibfnamefont {P.}~\bibnamefont {McGeehin}}, \bibinfo {author}
  {\bibfnamefont {M.}~\bibnamefont {Gubbins}}, \ and\ \bibinfo {author}
  {\bibfnamefont {V.}~\bibnamefont {Venugopal}},\ }\bibfield  {title} {\enquote
  {\bibinfo {title} {Whole wafer magnetostriction metrology for magnetic films
  and multilayers},}\ }\href@noop {} {\bibfield  {journal} {\bibinfo  {journal}
  {Measurement Science and Technology}\ }\textbf {\bibinfo {volume} {24}},\
  \bibinfo {pages} {045601} (\bibinfo {year} {2013})}\BibitemShut {NoStop}%
\bibitem [{\citenamefont {Raghunathan}, \citenamefont {Snyder},\ and\
  \citenamefont {Jiles}(2009)}]{raghunathan2009comparison}%
  \BibitemOpen
  \bibfield  {author} {\bibinfo {author} {\bibfnamefont {A.}~\bibnamefont
  {Raghunathan}}, \bibinfo {author} {\bibfnamefont {J.~E.}\ \bibnamefont
  {Snyder}}, \ and\ \bibinfo {author} {\bibfnamefont {D.}~\bibnamefont
  {Jiles}},\ }\bibfield  {title} {\enquote {\bibinfo {title} {Comparison of
  alternative techniques for characterizing magnetostriction and inverse
  magnetostriction in magnetic thin films},}\ }\href@noop {} {\bibfield
  {journal} {\bibinfo  {journal} {IEEE Transactions on Magnetics}\ }\textbf
  {\bibinfo {volume} {45}},\ \bibinfo {pages} {3269--3273} (\bibinfo {year}
  {2009})}\BibitemShut {NoStop}%
\bibitem [{\citenamefont {Cullity}\ and\ \citenamefont
  {Graham}(2011)}]{cullity2011introduction}%
  \BibitemOpen
  \bibfield  {author} {\bibinfo {author} {\bibfnamefont {B.~D.}\ \bibnamefont
  {Cullity}}\ and\ \bibinfo {author} {\bibfnamefont {C.~D.}\ \bibnamefont
  {Graham}},\ }\href@noop {} {\emph {\bibinfo {title} {Introduction to magnetic
  materials}}}\ (\bibinfo  {publisher} {John Wiley \& Sons},\ \bibinfo {year}
  {2011})\BibitemShut {NoStop}%
\bibitem [{\citenamefont {Zhang}\ \emph {et~al.}(2011)\citenamefont {Zhang},
  \citenamefont {Fan}, \citenamefont {Wang}, \citenamefont {Kou}, \citenamefont
  {Cao}, \citenamefont {Chen}, \citenamefont {{Ni}}, \citenamefont {Pan},\ and\
  \citenamefont {Xiao}}]{zhang2011study}%
  \BibitemOpen
  \bibfield  {author} {\bibinfo {author} {\bibfnamefont {Y.}~\bibnamefont
  {Zhang}}, \bibinfo {author} {\bibfnamefont {X.}~\bibnamefont {Fan}}, \bibinfo
  {author} {\bibfnamefont {W.}~\bibnamefont {Wang}}, \bibinfo {author}
  {\bibfnamefont {X.}~\bibnamefont {Kou}}, \bibinfo {author} {\bibfnamefont
  {R.}~\bibnamefont {Cao}}, \bibinfo {author} {\bibfnamefont {X.}~\bibnamefont
  {Chen}}, \bibinfo {author} {\bibfnamefont {C.}~\bibnamefont {{Ni}}}, \bibinfo
  {author} {\bibfnamefont {L.}~\bibnamefont {Pan}}, \ and\ \bibinfo {author}
  {\bibfnamefont {J.~Q.}\ \bibnamefont {Xiao}},\ }\bibfield  {title} {\enquote
  {\bibinfo {title} {Study and tailoring spin dynamic properties of {CoFeB}
  during rapid thermal annealing},}\ }\href@noop {} {\bibfield  {journal}
  {\bibinfo  {journal} {Applied Physics Letters}\ }\textbf {\bibinfo {volume}
  {98}},\ \bibinfo {pages} {042506} (\bibinfo {year} {2011})}\BibitemShut
  {NoStop}%
\bibitem [{\citenamefont {Thomas}\ \emph {et~al.}(2003)\citenamefont {Thomas},
  \citenamefont {Shen}, \citenamefont {Schieffer}, \citenamefont {Tournerie},\
  and\ \citenamefont {L{\'e}pine}}]{thomas2003interplay}%
  \BibitemOpen
  \bibfield  {author} {\bibinfo {author} {\bibfnamefont {O.}~\bibnamefont
  {Thomas}}, \bibinfo {author} {\bibfnamefont {Q.}~\bibnamefont {Shen}},
  \bibinfo {author} {\bibfnamefont {P.}~\bibnamefont {Schieffer}}, \bibinfo
  {author} {\bibfnamefont {N.}~\bibnamefont {Tournerie}}, \ and\ \bibinfo
  {author} {\bibfnamefont {B.}~\bibnamefont {L{\'e}pine}},\ }\bibfield  {title}
  {\enquote {\bibinfo {title} {Interplay between anisotropic strain relaxation
  and uniaxial interface magnetic anisotropy in epitaxial {Fe} films on (001)
  {GaAs}},}\ }\href@noop {} {\bibfield  {journal} {\bibinfo  {journal}
  {Physical Review Letters}\ }\textbf {\bibinfo {volume} {90}},\ \bibinfo
  {pages} {017205} (\bibinfo {year} {2003})}\BibitemShut {NoStop}%
\bibitem [{\citenamefont {McGrouther}\ \emph {et~al.}(2007)\citenamefont
  {McGrouther}, \citenamefont {McVitie}, \citenamefont {Chapman},\ and\
  \citenamefont {Gentils}}]{mcgrouther2007controlled}%
  \BibitemOpen
  \bibfield  {author} {\bibinfo {author} {\bibfnamefont {D.}~\bibnamefont
  {McGrouther}}, \bibinfo {author} {\bibfnamefont {S.}~\bibnamefont {McVitie}},
  \bibinfo {author} {\bibfnamefont {J.}~\bibnamefont {Chapman}}, \ and\
  \bibinfo {author} {\bibfnamefont {A.}~\bibnamefont {Gentils}},\ }\bibfield
  {title} {\enquote {\bibinfo {title} {Controlled domain wall injection into
  ferromagnetic nanowires from an optimized pad geometry},}\ }\href@noop {}
  {\bibfield  {journal} {\bibinfo  {journal} {Applied Physics Letters}\
  }\textbf {\bibinfo {volume} {91}},\ \bibinfo {pages} {022506} (\bibinfo
  {year} {2007})}\BibitemShut {NoStop}%
\bibitem [{\citenamefont {Wang}(2019)}]{wang2019mechanical}%
  \BibitemOpen
  \bibfield  {author} {\bibinfo {author} {\bibfnamefont {J.}~\bibnamefont
  {Wang}},\ }\bibfield  {title} {\enquote {\bibinfo {title} {Mechanical control
  of magnetic order: from phase transition to skyrmions},}\ }\href@noop {}
  {\bibfield  {journal} {\bibinfo  {journal} {Annual Review of Materials
  Research}\ }\textbf {\bibinfo {volume} {49}},\ \bibinfo {pages} {361--388}
  (\bibinfo {year} {2019})}\BibitemShut {NoStop}%
\bibitem [{\citenamefont {Bur}\ \emph {et~al.}(2011)\citenamefont {Bur},
  \citenamefont {Wu}, \citenamefont {Hockel}, \citenamefont {Hsu},
  \citenamefont {Kim}, \citenamefont {Chung}, \citenamefont {Wong},
  \citenamefont {Wang},\ and\ \citenamefont {Carman}}]{bur2011strain}%
  \BibitemOpen
  \bibfield  {author} {\bibinfo {author} {\bibfnamefont {A.}~\bibnamefont
  {Bur}}, \bibinfo {author} {\bibfnamefont {T.}~\bibnamefont {Wu}}, \bibinfo
  {author} {\bibfnamefont {J.}~\bibnamefont {Hockel}}, \bibinfo {author}
  {\bibfnamefont {C.-J.}\ \bibnamefont {Hsu}}, \bibinfo {author} {\bibfnamefont
  {H.~K.}\ \bibnamefont {Kim}}, \bibinfo {author} {\bibfnamefont {T.-K.}\
  \bibnamefont {Chung}}, \bibinfo {author} {\bibfnamefont {K.}~\bibnamefont
  {Wong}}, \bibinfo {author} {\bibfnamefont {K.~L.}\ \bibnamefont {Wang}}, \
  and\ \bibinfo {author} {\bibfnamefont {G.~P.}\ \bibnamefont {Carman}},\
  }\bibfield  {title} {\enquote {\bibinfo {title} {Strain-induced magnetization
  change in patterned ferromagnetic nickel nanostructures},}\ }\href@noop {}
  {\bibfield  {journal} {\bibinfo  {journal} {Journal of Applied Physics}\
  }\textbf {\bibinfo {volume} {109}},\ \bibinfo {pages} {123903} (\bibinfo
  {year} {2011})}\BibitemShut {NoStop}%
\bibitem [{\citenamefont {Im}\ \emph {et~al.}(2009)\citenamefont {Im},
  \citenamefont {Bocklage}, \citenamefont {Fischer},\ and\ \citenamefont
  {Meier}}]{im2009direct}%
  \BibitemOpen
  \bibfield  {author} {\bibinfo {author} {\bibfnamefont {M.-Y.}\ \bibnamefont
  {Im}}, \bibinfo {author} {\bibfnamefont {L.}~\bibnamefont {Bocklage}},
  \bibinfo {author} {\bibfnamefont {P.}~\bibnamefont {Fischer}}, \ and\
  \bibinfo {author} {\bibfnamefont {G.}~\bibnamefont {Meier}},\ }\bibfield
  {title} {\enquote {\bibinfo {title} {Direct observation of stochastic
  domain-wall depinning in magnetic nanowires},}\ }\href@noop {} {\bibfield
  {journal} {\bibinfo  {journal} {Physical Review Letters}\ }\textbf {\bibinfo
  {volume} {102}},\ \bibinfo {pages} {147204} (\bibinfo {year}
  {2009})}\BibitemShut {NoStop}%
\bibitem [{\citenamefont {Bogart}\ \emph {et~al.}(2009)\citenamefont {Bogart},
  \citenamefont {Atkinson}, \citenamefont {O’Shea}, \citenamefont
  {McGrouther},\ and\ \citenamefont {McVitie}}]{bogart2009dependence}%
  \BibitemOpen
  \bibfield  {author} {\bibinfo {author} {\bibfnamefont {L.}~\bibnamefont
  {Bogart}}, \bibinfo {author} {\bibfnamefont {D.}~\bibnamefont {Atkinson}},
  \bibinfo {author} {\bibfnamefont {K.}~\bibnamefont {O’Shea}}, \bibinfo
  {author} {\bibfnamefont {D.}~\bibnamefont {McGrouther}}, \ and\ \bibinfo
  {author} {\bibfnamefont {S.}~\bibnamefont {McVitie}},\ }\bibfield  {title}
  {\enquote {\bibinfo {title} {Dependence of domain wall pinning potential
  landscapes on domain wall chirality and pinning site geometry in planar
  nanowires},}\ }\href@noop {} {\bibfield  {journal} {\bibinfo  {journal}
  {Physical Review B}\ }\textbf {\bibinfo {volume} {79}},\ \bibinfo {pages}
  {054414} (\bibinfo {year} {2009})}\BibitemShut {NoStop}%
\bibitem [{\citenamefont {Backes}\ \emph {et~al.}(2007)\citenamefont {Backes},
  \citenamefont {Schieback}, \citenamefont {Kl{\"a}ui}, \citenamefont
  {Junginger}, \citenamefont {Ehrke}, \citenamefont {Nielaba}, \citenamefont
  {R{\"u}diger}, \citenamefont {Heyderman}, \citenamefont {Chen}, \citenamefont
  {Kasama} \emph {et~al.}}]{backes2007transverse}%
  \BibitemOpen
  \bibfield  {author} {\bibinfo {author} {\bibfnamefont {D.}~\bibnamefont
  {Backes}}, \bibinfo {author} {\bibfnamefont {C.}~\bibnamefont {Schieback}},
  \bibinfo {author} {\bibfnamefont {M.}~\bibnamefont {Kl{\"a}ui}}, \bibinfo
  {author} {\bibfnamefont {F.}~\bibnamefont {Junginger}}, \bibinfo {author}
  {\bibfnamefont {H.}~\bibnamefont {Ehrke}}, \bibinfo {author} {\bibfnamefont
  {P.}~\bibnamefont {Nielaba}}, \bibinfo {author} {\bibfnamefont
  {U.}~\bibnamefont {R{\"u}diger}}, \bibinfo {author} {\bibfnamefont {L.~J.}\
  \bibnamefont {Heyderman}}, \bibinfo {author} {\bibfnamefont {C.-S.}\
  \bibnamefont {Chen}}, \bibinfo {author} {\bibfnamefont {T.}~\bibnamefont
  {Kasama}},  \emph {et~al.},\ }\bibfield  {title} {\enquote {\bibinfo {title}
  {Transverse domain walls in nanoconstrictions},}\ }\href@noop {} {\bibfield
  {journal} {\bibinfo  {journal} {Applied Physics Letters}\ }\textbf {\bibinfo
  {volume} {91}},\ \bibinfo {pages} {112502} (\bibinfo {year}
  {2007})}\BibitemShut {NoStop}%
\bibitem [{\citenamefont {Weiler}\ \emph {et~al.}(2009)\citenamefont {Weiler},
  \citenamefont {Brandlmaier}, \citenamefont {Gepr{\"a}gs}, \citenamefont
  {Althammer}, \citenamefont {Opel}, \citenamefont {Bihler}, \citenamefont
  {Huebl}, \citenamefont {Brandt}, \citenamefont {Gross},\ and\ \citenamefont
  {G{\"o}nnenwein}}]{weiler2009voltage}%
  \BibitemOpen
  \bibfield  {author} {\bibinfo {author} {\bibfnamefont {M.}~\bibnamefont
  {Weiler}}, \bibinfo {author} {\bibfnamefont {A.}~\bibnamefont {Brandlmaier}},
  \bibinfo {author} {\bibfnamefont {S.}~\bibnamefont {Gepr{\"a}gs}}, \bibinfo
  {author} {\bibfnamefont {M.}~\bibnamefont {Althammer}}, \bibinfo {author}
  {\bibfnamefont {M.}~\bibnamefont {Opel}}, \bibinfo {author} {\bibfnamefont
  {C.}~\bibnamefont {Bihler}}, \bibinfo {author} {\bibfnamefont
  {H.}~\bibnamefont {Huebl}}, \bibinfo {author} {\bibfnamefont {M.~S.}\
  \bibnamefont {Brandt}}, \bibinfo {author} {\bibfnamefont {R.}~\bibnamefont
  {Gross}}, \ and\ \bibinfo {author} {\bibfnamefont {S.~T.}\ \bibnamefont
  {G{\"o}nnenwein}},\ }\bibfield  {title} {\enquote {\bibinfo {title} {Voltage
  controlled inversion of magnetic anisotropy in a ferromagnetic thin film at
  room temperature},}\ }\href@noop {} {\bibfield  {journal} {\bibinfo
  {journal} {New Journal of Physics}\ }\textbf {\bibinfo {volume} {11}},\
  \bibinfo {pages} {013021} (\bibinfo {year} {2009})}\BibitemShut {NoStop}%
\bibitem [{\citenamefont {Brandlmaier}\ \emph {et~al.}(2008)\citenamefont
  {Brandlmaier}, \citenamefont {Gepr{\"a}gs}, \citenamefont {Weiler},
  \citenamefont {Boger}, \citenamefont {Opel}, \citenamefont {Huebl},
  \citenamefont {Bihler}, \citenamefont {Brandt}, \citenamefont {Botters},
  \citenamefont {Grundler} \emph {et~al.}}]{brandlmaier2008situ}%
  \BibitemOpen
  \bibfield  {author} {\bibinfo {author} {\bibfnamefont {A.}~\bibnamefont
  {Brandlmaier}}, \bibinfo {author} {\bibfnamefont {S.}~\bibnamefont
  {Gepr{\"a}gs}}, \bibinfo {author} {\bibfnamefont {M.}~\bibnamefont {Weiler}},
  \bibinfo {author} {\bibfnamefont {A.}~\bibnamefont {Boger}}, \bibinfo
  {author} {\bibfnamefont {M.}~\bibnamefont {Opel}}, \bibinfo {author}
  {\bibfnamefont {H.}~\bibnamefont {Huebl}}, \bibinfo {author} {\bibfnamefont
  {C.}~\bibnamefont {Bihler}}, \bibinfo {author} {\bibfnamefont {M.~S.}\
  \bibnamefont {Brandt}}, \bibinfo {author} {\bibfnamefont {B.}~\bibnamefont
  {Botters}}, \bibinfo {author} {\bibfnamefont {D.}~\bibnamefont {Grundler}},
  \emph {et~al.},\ }\bibfield  {title} {\enquote {\bibinfo {title} {In situ
  manipulation of magnetic anisotropy in magnetite thin films},}\ }\href@noop
  {} {\bibfield  {journal} {\bibinfo  {journal} {Physical Review B}\ }\textbf
  {\bibinfo {volume} {77}},\ \bibinfo {pages} {104445} (\bibinfo {year}
  {2008})}\BibitemShut {NoStop}%
\bibitem [{\citenamefont {McMichael}\ and\ \citenamefont
  {Donahue}(1997)}]{mcmichael1997head}%
  \BibitemOpen
  \bibfield  {author} {\bibinfo {author} {\bibfnamefont {R.~D.}\ \bibnamefont
  {McMichael}}\ and\ \bibinfo {author} {\bibfnamefont {M.~J.}\ \bibnamefont
  {Donahue}},\ }\bibfield  {title} {\enquote {\bibinfo {title} {Head to head
  domain wall structures in thin magnetic strips},}\ }\href@noop {} {\bibfield
  {journal} {\bibinfo  {journal} {IEEE Transactions on Magnetics}\ }\textbf
  {\bibinfo {volume} {33}},\ \bibinfo {pages} {4167--4169} (\bibinfo {year}
  {1997})}\BibitemShut {NoStop}%
\bibitem [{\citenamefont {Vansteenkiste}\ \emph {et~al.}(2014)\citenamefont
  {Vansteenkiste}, \citenamefont {Leliaert}, \citenamefont {Dvornik},
  \citenamefont {Helsen}, \citenamefont {Garcia-Sanchez},\ and\ \citenamefont
  {Van~Waeyenberge}}]{vansteenkiste2014design}%
  \BibitemOpen
  \bibfield  {author} {\bibinfo {author} {\bibfnamefont {A.}~\bibnamefont
  {Vansteenkiste}}, \bibinfo {author} {\bibfnamefont {J.}~\bibnamefont
  {Leliaert}}, \bibinfo {author} {\bibfnamefont {M.}~\bibnamefont {Dvornik}},
  \bibinfo {author} {\bibfnamefont {M.}~\bibnamefont {Helsen}}, \bibinfo
  {author} {\bibfnamefont {F.}~\bibnamefont {Garcia-Sanchez}}, \ and\ \bibinfo
  {author} {\bibfnamefont {B.}~\bibnamefont {Van~Waeyenberge}},\ }\bibfield
  {title} {\enquote {\bibinfo {title} {The design and verification of
  {MuMax3}},}\ }\href@noop {} {\bibfield  {journal} {\bibinfo  {journal} {AIP
  Advances}\ }\textbf {\bibinfo {volume} {4}},\ \bibinfo {pages} {107133}
  (\bibinfo {year} {2014})}\BibitemShut {NoStop}%
\bibitem [{\citenamefont {Peng}\ \emph {et~al.}(2016)\citenamefont {Peng},
  \citenamefont {Hu}, \citenamefont {Momeni}, \citenamefont {Wang},
  \citenamefont {Chen},\ and\ \citenamefont {Nan}}]{peng2016fast}%
  \BibitemOpen
  \bibfield  {author} {\bibinfo {author} {\bibfnamefont {R.-C.}\ \bibnamefont
  {Peng}}, \bibinfo {author} {\bibfnamefont {J.-M.}\ \bibnamefont {Hu}},
  \bibinfo {author} {\bibfnamefont {K.}~\bibnamefont {Momeni}}, \bibinfo
  {author} {\bibfnamefont {J.-J.}\ \bibnamefont {Wang}}, \bibinfo {author}
  {\bibfnamefont {L.-Q.}\ \bibnamefont {Chen}}, \ and\ \bibinfo {author}
  {\bibfnamefont {C.-W.}\ \bibnamefont {Nan}},\ }\bibfield  {title} {\enquote
  {\bibinfo {title} {Fast 180° magnetization switching in a strain-mediated
  multiferroic heterostructure driven by a voltage},}\ }\href@noop {}
  {\bibfield  {journal} {\bibinfo  {journal} {Scientific Reports}\ }\textbf
  {\bibinfo {volume} {6}},\ \bibinfo {pages} {1--9} (\bibinfo {year}
  {2016})}\BibitemShut {NoStop}%
\bibitem [{\citenamefont {Yanes}\ \emph {et~al.}(2019)\citenamefont {Yanes},
  \citenamefont {Garcia-Sanchez}, \citenamefont {Luis}, \citenamefont
  {Martinez}, \citenamefont {Raposo}, \citenamefont {Torres},\ and\
  \citenamefont {Lopez-Diaz}}]{yanes2019skyrmion}%
  \BibitemOpen
  \bibfield  {author} {\bibinfo {author} {\bibfnamefont {R.}~\bibnamefont
  {Yanes}}, \bibinfo {author} {\bibfnamefont {F.}~\bibnamefont
  {Garcia-Sanchez}}, \bibinfo {author} {\bibfnamefont {R.}~\bibnamefont
  {Luis}}, \bibinfo {author} {\bibfnamefont {E.}~\bibnamefont {Martinez}},
  \bibinfo {author} {\bibfnamefont {V.}~\bibnamefont {Raposo}}, \bibinfo
  {author} {\bibfnamefont {L.}~\bibnamefont {Torres}}, \ and\ \bibinfo {author}
  {\bibfnamefont {L.}~\bibnamefont {Lopez-Diaz}},\ }\bibfield  {title}
  {\enquote {\bibinfo {title} {Skyrmion motion induced by voltage-controlled
  in-plane strain gradients},}\ }\href@noop {} {\bibfield  {journal} {\bibinfo
  {journal} {Applied Physics Letters}\ }\textbf {\bibinfo {volume} {115}},\
  \bibinfo {pages} {132401} (\bibinfo {year} {2019})}\BibitemShut {NoStop}%
\bibitem [{\citenamefont {Hubert}\ and\ \citenamefont
  {Sch{\"a}fer}(2008)}]{hubert2008magnetic}%
  \BibitemOpen
  \bibfield  {author} {\bibinfo {author} {\bibfnamefont {A.}~\bibnamefont
  {Hubert}}\ and\ \bibinfo {author} {\bibfnamefont {R.}~\bibnamefont
  {Sch{\"a}fer}},\ }\href@noop {} {\emph {\bibinfo {title} {Magnetic domains:
  the analysis of magnetic microstructures}}}\ (\bibinfo  {publisher} {Springer
  Science \& Business Media},\ \bibinfo {year} {2008})\BibitemShut {NoStop}%
\bibitem [{\citenamefont {Mollick}\ \emph {et~al.}(2018)\citenamefont
  {Mollick}, \citenamefont {Singh}, \citenamefont {Kumar}, \citenamefont
  {Bhattacharyya},\ and\ \citenamefont {Som}}]{mollick2018strong}%
  \BibitemOpen
  \bibfield  {author} {\bibinfo {author} {\bibfnamefont {S.~A.}\ \bibnamefont
  {Mollick}}, \bibinfo {author} {\bibfnamefont {R.}~\bibnamefont {Singh}},
  \bibinfo {author} {\bibfnamefont {M.}~\bibnamefont {Kumar}}, \bibinfo
  {author} {\bibfnamefont {S.}~\bibnamefont {Bhattacharyya}}, \ and\ \bibinfo
  {author} {\bibfnamefont {T.}~\bibnamefont {Som}},\ }\bibfield  {title}
  {\enquote {\bibinfo {title} {Strong uniaxial magnetic anisotropy in {Co}
  films on highly ordered grating-like nanopatterned {Ge} surfaces},}\
  }\href@noop {} {\bibfield  {journal} {\bibinfo  {journal} {Nanotechnology}\
  }\textbf {\bibinfo {volume} {29}},\ \bibinfo {pages} {125302} (\bibinfo
  {year} {2018})}\BibitemShut {NoStop}%
\bibitem [{\citenamefont {Kumar}\ \emph {et~al.}(2006)\citenamefont {Kumar},
  \citenamefont {F{\"a}hler}, \citenamefont {Schloerb}, \citenamefont
  {Leistner},\ and\ \citenamefont {Schultz}}]{kumar2006competition}%
  \BibitemOpen
  \bibfield  {author} {\bibinfo {author} {\bibfnamefont {A.}~\bibnamefont
  {Kumar}}, \bibinfo {author} {\bibfnamefont {S.}~\bibnamefont {F{\"a}hler}},
  \bibinfo {author} {\bibfnamefont {H.}~\bibnamefont {Schloerb}}, \bibinfo
  {author} {\bibfnamefont {K.}~\bibnamefont {Leistner}}, \ and\ \bibinfo
  {author} {\bibfnamefont {L.}~\bibnamefont {Schultz}},\ }\bibfield  {title}
  {\enquote {\bibinfo {title} {Competition between shape anisotropy and
  magnetoelastic anisotropy in {Ni} nanowires electrodeposited within alumina
  templates},}\ }\href@noop {} {\bibfield  {journal} {\bibinfo  {journal}
  {Physical Review B}\ }\textbf {\bibinfo {volume} {73}},\ \bibinfo {pages}
  {064421} (\bibinfo {year} {2006})}\BibitemShut {NoStop}%
\bibitem [{\citenamefont {Gabor}\ \emph {et~al.}(2011)\citenamefont {Gabor},
  \citenamefont {Petrisor~Jr}, \citenamefont {Tiusan}, \citenamefont {Hehn},\
  and\ \citenamefont {Petrisor}}]{gabor2011magnetic}%
  \BibitemOpen
  \bibfield  {author} {\bibinfo {author} {\bibfnamefont {M.}~\bibnamefont
  {Gabor}}, \bibinfo {author} {\bibfnamefont {T.}~\bibnamefont {Petrisor~Jr}},
  \bibinfo {author} {\bibfnamefont {C.}~\bibnamefont {Tiusan}}, \bibinfo
  {author} {\bibfnamefont {M.}~\bibnamefont {Hehn}}, \ and\ \bibinfo {author}
  {\bibfnamefont {T.}~\bibnamefont {Petrisor}},\ }\bibfield  {title} {\enquote
  {\bibinfo {title} {Magnetic and structural anisotropies of heusler alloy
  epitaxial thin films},}\ }\href@noop {} {\bibfield  {journal} {\bibinfo
  {journal} {Physical Review B}\ }\textbf {\bibinfo {volume} {84}},\ \bibinfo
  {pages} {134413} (\bibinfo {year} {2011})}\BibitemShut {NoStop}%
\end{thebibliography}%

\newpage



\end{document}